\documentclass[12pt]{article}
\usepackage{amsmath}
\usepackage{amssymb}
\usepackage{graphicx}

\numberwithin{equation}{section}

\newcommand{\Tr}{\mbox{\rm Tr}}
\begin{document}
\baselineskip=18pt
\begin{titlepage}
\begin{flushright}
{\small KYUSHU-HET-110}\\%
{\small TU-808}\\%
{\small OU-HET 595/2008}%
\end{flushright}
\begin{center}
\vspace*{11mm}

{\large\bf%
Multi-Higgs Mass Spectrum in Gauge-Higgs Unification%
}\vspace*{8mm}

Kentaro Kojima$^{a,}$\footnote{E-mail: kojima@higgs.phys.kyushu-u.ac.jp}, 
Kazunori Takenaga$^{b,}$\footnote{E-mail: takenaga@tuhep.phys.tohoku.ac.jp}, 
and 
Toshifumi Yamashita$^{c,}$\footnote{E-mail: yamasita@het.phys.osaka-u.ac.jp}%
\vspace*{5mm}

{\it $^a$ Department of Physics, Kyushu University, Fukuoka 812-8581, Japan\\
$^b$ Department of Physics, Tohoku University, Sendai 980-8578, Japan\\
$^c$ Department of Physics, Osaka University, Toyonaka, Osaka 560-0043, Japan%
}

\vspace*{3mm}

{\small (January, 2008)}%
\end{center}
\vspace*{5mm}

\begin{abstract}\noindent%
We study an $SU(2)$ supersymmetric gauge model in a framework of 
gauge-Higgs unification. Multi-Higgs spectrum appears in the model at 
low energy. We develop a useful perturbative approximation scheme for 
evaluating effective potential to study the multi-Higgs mass spectrum. 
We find that both tree-massless and massive Higgs scalars obtain mass 
corrections of similar size from finite parts of the loop effects. 
The corrections modify multi-Higgs mass spectrum, 
and hence, the loop effects are significant in view of future verifications 
of the gauge-Higgs unification scenario in high-energy experiments. 
\end{abstract}
\end{titlepage}
\newpage
\section{Introduction}
\label{sec:int}
The standard model (SM) of particle physics successfully describes 
high-energy experimental data. However, the SM involves the theoretical 
difficulty referred to as ``hierarchy problem'': due to large quantum 
corrections, the energy scale of the Higgs potential tends to become similar 
to the cutoff scale of the SM, in spite of the fact that the energy scale 
should be similar to the known electroweak one. This suggests that physics 
beyond the SM appears near the electroweak scale. 

Among several candidates of the physics beyond the SM, unification 
of gauge and Higgs fields, so-called gauge-Higgs unification, is known as 
a promising idea with the help of compactified extra 
dimensions~\cite{manton,fair,hosotani}. In the gauge-Higgs unification 
scenario, some of the extra-dimensional components of gauge fields are 
identified to Higgs fields at a low-energy regime. Phenomenologically 
viable electroweak Higgs sectors are known to appear with orbifold 
compactification of the extra dimensions. An advantageous point of the 
idea is finiteness of the Higgs potential even at some loop levels; hence, 
the potential is not sensitive to unknown ultra-violet (UV) physics, and 
the scenario gives a reasonable resolution to the hierarchy problem. 
The gauge-Higgs unification scenario has been extensively 
studied~\cite{gaugehiggs1,gaugehiggs2,gaugehiggs3,models}. In addition to 
the case with flat compactified extra dimensions, the case with a warped 
extra dimension has been also investigated~\cite{warp}. The Higgs sector is 
predictive thanks to the higher dimensional gauge invariance. Namely, 
the scenario predicts light Higgs scalars, which can be a key ingredient 
for the experimental test of the scenario at LHC and ILC.

One of the most important subjects is to minutely examine the Higgs mass 
spectrum in the gauge-Higgs unification. Several models have been 
proposed and the mass spectrum has been studied. An interesting observation 
is that multi-Higgs scalars appear in some models. The models lead to 
non-vanishing tree-level Higgs potential which has flat directions; 
it has revealed that finite radiative corrections to the flat direction lead 
to the correct electroweak symmetry breaking 
dynamically~\cite{multi6d,multi5d}. The Higgs scalars associated with the 
flat directions are massless at the tree-level and become massive through 
the radiative corrections; their masses have been studied by the one-loop 
effective potential~\cite{higgsmass,higgsmass1}, or even at the two-loop 
level in a simple model~\cite{tlghu}. 
The other modes among the multi-Higgs scalars, 
which are associated with the non-flat directions, are 
massive at the tree-level. Loop corrections to their masses, however, have not 
been focused on even though they are crucial to verify the Higgs mass spectrum 
in the models. It is also important to examine the spectrum 
in detail for the experimental test of the scenario. 

In addition, 
the loop corrections to the non-flat directions are considered to be 
significant for determining the vacuum structure of the multi-Higgs potential. 
If the corrections are taken into account, then vacua deviated from the flat 
directions of the tree-level potential may appear. Clearly, such a vacuum 
structure cannot be studied when one focus only on the loop corrections 
to the flat directions. 
Thus, to reveal the correct vacuum structure of multi-Higgs 
models, the loop corrections to the non-flat directions should be 
incorporated. 

In this paper, we study the multi-Higgs mass spectrum based 
on one-loop effective potential in a simple 5D model. We take bulk loop 
corrections into account for all the modes of the Higgs scalars: we 
study not only the loop corrections for the Higgs scalars along the 
flat direction of the tree-level potential, but also the ones for 
the Higgs scalars along the non-flat direction. As mentioned above, 
in the past studies, the latter has not been focused on though it 
is important. In our analyses, it is found that the one-loop corrections 
involve an UV divergence, which is proportional to the tree-level potential 
and is renormalized. 
The one-loop corrected multi-Higgs mass spectrum is derived 
through the effective potential after eliminating the divergence. It turns 
out that both the tree-massless and massive Higgs scalars have mass 
corrections of similar size from finite parts of the one-loop effects. 
Consequently, the loop effects modify multi-Higgs mass spectrum and are 
significant in view of verifications of the scenario in 
high-energy experiments. 

The outline of the paper is as follows. In Section~\ref{sec:mass}, 
an overview of multi-Higgs mass spectrum in the gauge-Higgs unification 
scenario is presented. In Section~\ref{sec:5dsusy}, we examine 
the one-loop corrected multi-Higgs mass spectrum in a simple 5D model. 
A perturbative calculation of the effective potential is developed to 
estimate the loop corrections. Summary and future perspective 
are given in Section~\ref{sec:sum}. In Appendix~\ref{sec:op}, 
field dependent operators that are needed to evaluate the effective 
potential in the model are presented, and Appendix~\ref{sec:loop} provides 
the evaluation of loop momentum integrals with the summation of 
Kaluza-Klein (K-K) modes. 

\section{An overview of multi-Higgs mass spectrum}
\label{sec:mass} 
In this section, an overview of the multi-Higgs mass spectrum is given. 
An interesting observation is that multi-Higgs scalars are predicted in 
some models of the gauge-Higgs unification. Models with two extra dimensions 
compactified on the orbifold $T^2/Z_2$ are illustrative examples; at a 
low-energy regime, the two extra dimensional components of a 6D gauge field 
behave as a pair of scalar fields, those are identified as two Higgs 
scalars~\cite{multi6d}. It is also known that there appear multi-Higgs 
scalars in the 5D models with supersymmetry (SUSY)~\cite{multi5d,correcthp}. 
Due to the SUSY, a real scalar field is accompanied by a extra dimensional 
component of 5D gauge field (and a Weyl fermion) to form 4D ${\cal N}=1$ 
chiral superfield~\cite{nima}; two Higgs scalars hence appear at low energy. 

In the above models, the low-energy effective theory has a tree-level 
Higgs potential, where flat directions appear. Among the Higgs scalars, 
a mode, which corresponds to a flat direction and is a part of the zero modes 
of the extra dimensional component of the gauge field, 
triggers off the electroweak 
symmetry breaking. Quantum corrections to the flat direction determine 
the physical vacuum of the theory in terms of the non-zero vacuum expectation 
value (VEV) of the Wilson line phase degrees of freedom. Such the dynamical 
gauge symmetry breaking is known as the Hosotani mechanism~\cite{hosotani}.

We refer to the Higgs scalar that is the relevant mode to the symmetry 
breaking as the Symmetry-Breaking (SB) Higgs hereafter. The SB Higgs 
is identified as a zero-mode of extra dimensional components of gauge fields. 
In addition, gauge fields, non-SB Higgs scalars and K-K modes generally 
appear. The one-loop effective potential that takes account of only the 
SB Higgs background has been studied; 
the correct dynamical electroweak symmetry 
breaking has been shown to occur through the VEV of the SB 
Higgs~\cite{multi6d,multi5d}.  

After the symmetry breaking, there appear the massive gauge fields, whose 
mass scale $M_W$ should be much smaller than the typical mass scale of 
the K-K modes, ${\cal O}(1/R)$, where $R$ is the radius of a compactified 
dimension. 
In the 5D case, for instance, the mass of the gauge boson is typically 
given by $\alpha/R$, where $\alpha$ is the VEV of 
the Wilson line phase degree 
of freedom. The phase is related with the SB Higgs as 
$\alpha=2\pi g R\langle{a}\rangle$ (mod $2\pi$), where $g$ is the 5D gauge 
coupling and $a$ is a classical background of the extra dimensional component 
of the gauge fields, namely the SB Higgs. Some of the non-SB Higgs scalars 
also have interaction between the SB Higgs in the tree-level potential 
and have masses of ${\cal O}(\alpha/R)$. Since the K-K modes should 
be sufficiently heavy, a realistic vacuum satisfies $\alpha\ll 1$ and such 
a vacuum is dynamically realized with an appropriate choice of the bulk matter 
fields as discussed in the literature~\cite{multi5d}. 

The SB Higgs has no mass term in the tree-level potential; the mass 
arises from the one-loop effective potential. 
The SB Higgs tends to be lighter 
than the massive gauge fields due to the loop suppression factor, 
which is inconsistent 
with the experimental bounds of Higgs searches~\cite{PDG07}. 
A phenomenologically viable SB Higgs mass, however, can be realized through 
enhancement mechanism, as shown in~\cite{higgsmass,higgsmass1,csaba}.

The loop corrections to the masses of the non-SB Higgs have not been 
frequently discussed in the gauge-Higgs unification. The large one-loop 
mass correction to the SB Higgs mass implies that there also 
appear ${\cal O}(M_W)$ or larger one-loop contributions to the non-SB 
Higgs masses. The corrections are not suppressed rather than the tree-level 
masses of the non-SB Higgs, and thus are expected to give important effects on
the multi-Higgs mass spectrum. In order to see this explicitly, 
we examine the one-loop corrections and multi-Higgs mass spectrum in 
a simple model. 

\section{Multi-Higgs mass spectrum in a 5D SUSY model}
\label{sec:5dsusy}
\subsection{Setup}
In this section, we consider a 5D SUSY gauge theory, where the 
fifth dimension is assumed to be compactified on the orbifold $S^1/Z_2$ 
and a multi-Higgs spectrum at low energy is predicted. 
On the orbifold, a point of the fifth dimensional coordinate $y$ is 
identified with other points by the translation, $U[y+2\pi R]\sim y$, 
and the reflection, $P_0[-y]\sim y$. Combining them, one can define 
another reflection operator $P_1\equiv U^{-1}P_0$, then $\pi R-y$ and 
$\pi R+y$ are identified by the reflection $P_1[\pi R-y]=\pi R+y$. 
Since the points $y=0,\ \pi R$ are invariant under the reflections 
respectively, they 
are called orbifold fixed points. 

The vector multiplet ${\cal V}$ in the 5D SUSY theory can be decomposed into a 
4D chiral superfield $\Phi$ and a real vector superfield $V$ as follows:
\begin{eqnarray}
  {\cal V}&=&(A_M^a,\eta_1^a,\eta_2^a,\Sigma^a)\;\to \;
  \begin{cases}
    V\;=\;(A_\mu^a,\eta_1^a),\\
    \Phi\;=\;(\Sigma^a+iA_y^a,\eta_2^a),
  \end{cases}
\label{vecmul}
\end{eqnarray}
where $M=(\mu,y)$ and $A_M^a$, $\eta_{1,2}^a$ and $\Sigma^a$ are the 5D gauge 
field, Majorana spinors and a real scalar, respectively~\cite{nima}. 
The subscript $a$ denotes the index of 
the adjoint representation of gauge group. 
The Lagrangian of the 5D SUSY theory is given as follows~\cite{sohnius}: 
\begin{eqnarray}
  {\cal L}_{\rm vec}&=&\Tr\left[-{1\over 2}F_{MN}F^{MN}+(D_M\Sigma)D^M\Sigma
    +\bar \lambda_i i\Gamma^MD_M\lambda_i-g\bar \lambda_i[\Sigma,\lambda_i]
    \right]
\label{lagvec},
\end{eqnarray}
where $g$ is the five dimensional gauge coupling constant and the Gamma 
matrices are defined by $(\Gamma^\mu,\Gamma^y)=(\gamma^\mu,i\gamma^5)$. 
The field strength and the covariant derivatives are defined by
\begin{eqnarray}
  F_{MN}&\equiv &\partial_MA_N-\partial_NA_M-ig[A_M,A_N],\qquad 
D_M\phi\;\equiv\; \partial_M\phi-ig[A_M,\phi],
\end{eqnarray}
where $\phi$ implies fields in the adjoint representation. Fermions 
$\lambda_i$ $(i=1,2)$ are symplectic-Majorana spinors; they are written by 
\begin{eqnarray}
  \lambda_i^a&=&
  \begin{pmatrix}
    \eta_i^a\\
    \epsilon_{ij}(i\sigma^2)\eta_{j}^{a*}
  \end{pmatrix},
\end{eqnarray}
where $\epsilon_{ij}$ is antisymmetric with $\epsilon_{12}=1$ and 
$\sigma^2$ is the Pauli matrix 
in the spinor space. It is known that the Lagrangian~\eqref{lagvec} has a 
global symmetry, called $SU(2)_R$, and the symplectic-Majorana 
spinors are transformed as doublets under the symmetry. 

One can introduce bulk hypermultiplets in the theory. The Lagrangian is 
written as follows~\cite{sohnius}:
\begin{eqnarray}\notag
  {\cal L}_{\rm mat}&=&|D_M\phi_i|^2
-g^2\phi^\dag_i\Sigma^2\phi_i-{g^2\over 2}\sum_{a,m}\left(
\phi_i^\dag (\tau^m)_{ij}t^a_\phi \phi_j\right)^2\\
&&+\bar {\psi}(i\Gamma^MD_M-g\Sigma)\psi
-(ig\sqrt{2}\bar  {\psi}\phi_i \lambda_i+{\rm h.c.}),\\[2mm]
(D_M\phi)^\alpha&=&(\partial_M\delta^{\alpha\beta}
-igA_M^a(t^a_\phi)^{\alpha\beta})\phi^\beta,
\label{lagmat}
\end{eqnarray}
where $\phi_i$ ($i=1,2$) are complex scalars, 
$\psi=(\tilde \phi_L,\tilde \phi_R)^T$ is a Dirac spinor, $t^a_\phi$ is 
the representation matrix of $\phi$, and $\tau^m$ $(m=1,2,3)$ are $SU(2)_R$ 
generators.

The geometry $S^1/Z_2$ requires us to choose the boundary conditions for the 
fields: each field can have non-trivial transformations under the 
translation $U$ and the reflections $P_0$ and $P_1$ in such a way that the 
Lagrangian is invariant. Let $\varphi(x,y)$ be a general field in a 
representation space of the symmetry of theory. 
Transformation law of the field 
is defined by 
\begin{eqnarray}
  {\cal L}[\varphi(x,{\cal T}_i[y])]&\equiv&
  {\cal L}[{\cal U}_\varphi[{\cal T}_i]\varphi(x,y)],
\end{eqnarray}
where ${\cal T}_i=\{U,P_0,P_1\}$. The operator ${\cal U}_\varphi[{\cal T}_i]$
acts on the field in its representation space. The transformations of 
the coordinate satisfy 
\begin{eqnarray}
  P_0^2\;=\;P_1^2\;=\;{\bf 1}, \qquad UP_0U\;=\;P_0, \qquad P_1\;=\;U^{-1}P_0,
\label{ytrans}
\end{eqnarray}
where ${\bf 1}$ denotes the identity operation. Corresponding 
to~\eqref{ytrans}, transformations of the field should be chosen to 
satisfy the following set of constraints in order to keep the consistency 
of the translation and parity operations: 
\begin{eqnarray}
  {\cal U}_\varphi[P_0]^2\;=\;
  {\cal U}_\varphi[P_1]^2\;=\;{\bf 1}, 
  \quad 
  {\cal U}_\varphi[U]{\cal U}_\varphi[P_0]{\cal U}_\varphi[U]\;=\;
  {\cal U}_\varphi[P_0], 
  \quad 
  {\cal U}_\varphi[P_1]\;=\;{\cal U}_\varphi[U]^{-1}{\cal U}_\varphi[P_0].
\label{bcconst}
\end{eqnarray}
The transformation law is referred to as the boundary condition of each 
field~\cite{bcs}. Using the last equality of~\eqref{bcconst}, one can read 
the boundary condition of the parity $P_1$ from those of $U$ and $P_0$. 
The gauge symmetry of the theory can be broken through non-trivial boundary 
conditions. It is also known that the remaining ${\cal N}=1$ SUSY is 
explicitly broken with twisted boundary conditions for the $SU(2)_R$ doublets 
and we adopt this mechanism, 
the so-called Scherk-Schwarz SUSY breaking~\cite{SS}. 

In the following, we examine a simple toy model where the multi-Higgs scalars 
appear at low energy. The original gauge symmetry of the theory is assumed 
to be $SU(2)$; the symmetry is explicitly broken to $U(1)$ by the particular 
boundary conditions which satisfy~\eqref{bcconst}: 
for the 5D vector multiplet, we take
{\allowdisplaybreaks
\begin{eqnarray}\notag
  A_\mu(y+2\pi R)&=&A_\mu(y),\qquad   A_\mu(-y)\;=\;\tau_3 A_\mu(y)\tau_3,\\
\notag
  A_y(y+2\pi R)&=&A_y(y),\qquad   A_y(-y)\;=\;-\tau_3A_y(y)\tau_3,\\
\notag
  \Sigma (y+2\pi R)&=&\Sigma (y), \qquad   \quad \Sigma(-y)\;=\;
-\tau_3\Sigma(y)\tau_3,\\[2mm]\notag
  \begin{pmatrix}
    \eta_1\\
    \eta_2
  \end{pmatrix}(y+2\pi R)
&=&\begin{pmatrix}
  \cos{({2\pi \beta })}&-\sin{({2\pi\beta })}\\
  \sin{({2 \pi\beta })}&\cos{({2\pi\beta })}
\end{pmatrix}
  \begin{pmatrix}
    \eta_1\\
    \eta_2
  \end{pmatrix}(y),\\
  \begin{pmatrix}
    \eta_1\\
    \eta_2
  \end{pmatrix}(-y)
&=&
  \begin{pmatrix}
   \tau_3 \eta_1\tau_3\\
   -\tau_3 \eta_2\tau_3
  \end{pmatrix}(y),
\label{vecbc}
\end{eqnarray}
and for the hypermultiplets, we take
\begin{eqnarray}\notag
  \begin{pmatrix}
    \phi_1\\ \phi_2
  \end{pmatrix}(y+2\pi R)&=&
  \eta_U  \begin{pmatrix}
    \cos{({2\pi \beta })}&-\sin{({2\pi\beta })}\\
    \sin{({2 \pi\beta })}&\cos{({2\pi\beta })}
  \end{pmatrix}
  \begin{pmatrix}
    \phi_1 \\ 
    \phi_2 
  \end{pmatrix}(y),\\\notag
  \begin{pmatrix}
    \phi_1\\ \phi_2
  \end{pmatrix}(-y)&=&
  \eta_P  
  \begin{pmatrix}
    {\cal T}_\phi[t^3]    \phi_1 \\ 
    -{\cal T}_\phi[t^3] \phi_2 
  \end{pmatrix}(y),\\\notag
  \begin{pmatrix}
    \tilde \phi_L\\ \tilde \phi_R
  \end{pmatrix}(y+2\pi R)&=&
  \eta_U  
  \begin{pmatrix}
    \tilde \phi_L\\ 
    \tilde \phi_R
  \end{pmatrix}(y),\\
  \begin{pmatrix}
    \tilde \phi_L\\ \tilde \phi_R
  \end{pmatrix}(-y)&=&
  \eta_P
  \begin{pmatrix}
    {\cal T}_\phi[t^3]\tilde \phi_L\\ 
    -{\cal T}_\phi[t^3]\tilde \phi_R
  \end{pmatrix}(y),
  \label{hypbc}
\end{eqnarray}}\noindent
where $\tau_3$ (${\cal T}_\phi[t^3]$) is the diagonal generator of the 
$SU(2)$ gauge symmetry in the fundamental representation (in the 
representation of $\phi$). Additional parities of each hypermultiplet are 
incorporated by $\eta_U$ and $\eta_P$; they must be $1$ or $-1$ in order 
to satisfy the consistency conditions~\eqref{bcconst}. As mentioned above, 
we introduce the Scherk-Schwarz SUSY breaking with a parameter $\beta$ in 
a general form: if $\sin(2\pi\beta)\neq 0$, then the residual ${\cal N}=1$ 
SUSY is broken and there appears ${\cal O}(\beta/R)$ (mod $1/R$) mass 
splitting between bosonic and fermionic states in the theory~\cite{SS2}. 
With the boundary conditions~\eqref{vecbc} and~\eqref{hypbc}, among the 
gauge bosons only $A_\mu^3$ has massless zero mode, 
and it corresponds to the 
4D gauge boson of the residual $U(1)$ gauge symmetry. 

There appear four real scalar zero-modes, that is, Higgs scalars from the 5D 
vector multiplet~\eqref{vecmul} under the boundary conditions~\eqref{vecbc}. 
We regard them as classical backgrounds of the theory and take the 
substitution in the Lagrangian~\eqref{lagvec} and~\eqref{lagmat}: 
\begin{eqnarray}
  A_y^{a=1,2}(x^M)\;\to\; A_y^{a=1,2}(x^M)+a_{a=1,2},\qquad 
\Sigma^{a=1,2}(x^M)\;\to\; \Sigma^{a=1,2}(x^M)+\sigma_{a=1,2},
\label{sub}
\end{eqnarray}
where $a_a$ and $\sigma_a$ are the classical backgrounds. Except for the 
backgrounds, any fields in the theory are referred to as fluctuations. 
Among the backgrounds, $a_1$ and $a_2$ have Wilson line phase degrees of 
freedom and evolve non-trivial VEVs through quantum 
corrections~\cite{correcthp}. Using the residual $U(1)$ gauge transformation, 
one can freely rotate the direction of the VEVs in the field space spanned by 
$a_1$ and $a_2$. We take $a_2$ as the SB Higgs field in the analyses. 
Then, $a_1$ is eaten by the longitudinal mode of the zero-mode of 
the $U(1)$ gauge field after the symmetry breaking. The other physical 
Higgs modes, $\sigma_1$ and $\sigma_2$, are the non-SB Higgs. 

With the substitution~\eqref{sub} in the Lagrangian~\eqref{lagvec}, the square 
of the fifth dimensional covariant derivative of $\Sigma$ yields the classical 
potential of the theory: 
\begin{eqnarray}\notag
  V_{\rm tree}
  &=&{g^2\over 2}\sum_{a,a',c,c'=1}^2
\sum_{d=1}^3a_aa_{a'}
\sigma_c\sigma_{c'}f^{acd}f^{a'c'd}\\
&=&{g^2\over 2}(-\sigma_2a_1+a_2\sigma_1)^2, 
\label{treepotas}
\end{eqnarray}
where $f^{abc}$ is the structure constant of the $SU(2)$ gauge symmetry. 
If the SB-Higgs is expanded around a VEV as 
$a_2=\tilde a_2+\langle{a_2}\rangle$, then the potential involves 
mass term for $\sigma_1$, which corresponds to the Higgs scalar associated 
with the non-flat direction of the potential. As mentioned above, $a_1$ 
is eaten by the longitudinal mode of the residual $U(1)$ gauge boson. 
The other backgrounds, $\tilde a_2$ and $\sigma_2$, are massless at 
the tree-level. 

The backgrounds in~\eqref{sub} mix with each other through the residual 
$U(1)$ gauge transformation. It is useful to turn to a new basis where 
Higgs fields are eigenstates of the $U(1)$ gauge symmetry; 
with the following reparametrization, the classical backgrounds form a pair 
of complex scalars: 
\begin{eqnarray}
  n_u&=&{1\over 2}(ia_1+a_2+\sigma_1-i\sigma_2),\qquad 
  n_d\;=\;{1\over 2}(-ia_1+a_2-\sigma_1-i\sigma_2),
\end{eqnarray}
where they have opposite charge of the $U(1)$ gauge symmetry. 
We refer to $n_{u,d}$ as up- and down-type Higgs scalars. In this basis, 
the tree-level potential~\eqref{treepotas} takes a clearer form as 
\begin{eqnarray}\notag
  V_{\rm tree}&=& {g^2\over 2}\left(|n_u|^2-|n_d|^2\right)^2 \\
  &=&  {g^2\over 2}D^2,
  \label{treesu2susy}\\
  D&\equiv &-\sigma_2a_1+a_2\sigma_1\;=\;|n_u|^2-|n_d|^2. \label{dterm}
\end{eqnarray}
The form of the tree-level potential is constrained by the symmetries of 
the theory: it corresponds to the D-term potential of the residual $U(1)$ 
gauge symmetry in terms of ${\cal N}=1$ SUSY theory~\cite{nima}. The flat 
directions lie along $D=0$. If a linear term of $D$ is incorporated in 
the theory, non-zero VEV of the D-term is realized and the SUSY is 
spontaneously broken~\cite{FIterm}. 

\subsection{Perturbative evaluation of the effective potential}
In this subsection, we estimate the one-loop effective potential including 
all the modes of the Higgs scalars. The calculation is carried out with a 
straightforward way; at first we adopt K-K mode decomposition of all the 
fluctuations, and the fifth-dimensional coordinate is integrated out in the 
action. Then, we obtain the effective 4D theory and can estimate 
the functional integral with infinite towers of the K-K modes. As a result, 
contributions to the effective potential generally written by 
\begin{eqnarray}\notag
  \delta V&=&{-i\over 2\pi R}{N_B\over 2}
  \sum_{\rm K-K}
  \ln\det\left[
    \Delta_{0B}+m_B\Delta_{1B}+\Delta_{2B}
  \right]\\
  &&+{i\over 2\pi R}{N_F\over 2}
  \sum_{\rm K-K}
  \ln\det\left[\Delta_{0F}
    +m_F\Delta_{1F}+\Delta_{2F}
  \right],\label{veffgen}
\end{eqnarray}
where the determinants are taken over the 4D momentum space and representation 
space of the $SU(2)$ gauge symmetry. The first (second) line corresponds to 
the contributions from bosonic (fermionic) fluctuations. In the determinants, 
we take $\Delta_{0B,F}=(\square +m_{B,F}^2)\cdot {\bf 1}$ such that 
$m_{B,F}^2$ are independent of both the gauge coupling and the classical 
backgrounds, where {\bf 1} means identity matrix in representation space of 
the fluctuation fields in internal loops. The operators, $\Delta_{1}$ and 
$\Delta_2$, are of order $g^1$ and $g^2$, respectively. They also take forms 
of matrices in specific representation space. The overall factor, $N_{B(F)}$, 
counts the bosonic (fermionic) degree of freedom in the internal loops. 

If the eigenvalues of the operators are analytically obtained, then the 
functional determinant and the K-K mode summation may be directly evaluated. 
This is actually the case where one focuses only on a particular background 
of Wilson line phase degrees of freedom and the other backgrounds are set to 
zero. When one chooses $a_2\neq 0$ and $a_1=\sigma_1=\sigma_2=0$ for instance, 
a part of the contribution typically evaluated as 
follows~\cite{multi5d,higgsmass,correcthp,csaba,potentialSS1,potfor}:
\begin{eqnarray}
  {1\over 2\pi R}\int {d^4p_E\over (2\pi)^4}
  \sum_{n=-\infty}^\infty\ln\left[p_E^2+\left({n+gRa_2\over R}\right)^2\right]
  &=&{3\over  64\pi^7R^5}
  \sum_{w=1}^\infty{\cos(2\pi gRw a_2)\over w^5},
\end{eqnarray}
where $p_E$ is the Wick-rotated momentum and the background independent term 
is discarded in the right-hand side. The Wilson line phase degree of freedom 
$\alpha \equiv 2\pi g R a_2$ (mod $2\pi $) appears in the cosine function. 
For a small value of $\alpha\ll 1$, K-K modes are sufficiently heavier than 
the weak scale, and one can approximately evaluate the summation as 
\begin{eqnarray}
  \sum_{w=1}^\infty{\cos(\alpha w)\over w^5}
&\simeq&\zeta(5)-{\zeta(3)\over 2}\alpha^2+
{\alpha^4\over 288}\left(25-6\log[\alpha^2]\right)+{\cal O}(\alpha^6),
\label{perteg}
\end{eqnarray}
where $\zeta(s)$ is the Riemann zeta-function. The approximate expansion 
have a logarithmic singularity with $\alpha\to 0$; it implies the infra-red 
(IR) divergence of the zero-mode propagator in internal loops. A finite VEV 
of $\alpha$ gives non-zero masses of the zero-modes and thus the singularity 
disappears. 

When one includes the general background fields~\eqref{sub}, it is difficult 
to obtain all the eigenvalues and/or to carry out the summation as above. 
Here, we evaluate approximate forms of contributions to the effective 
potential as in~\eqref{perteg}. Functional determinants in~\eqref{veffgen} 
have a perturbative expansion  of the gauge coupling and the background fields:
\begin{eqnarray}
  \ln\det[\Delta_0+m\Delta_1+\Delta_2]
&=&\ln\det[\Delta_0]
+ \sum_{f=1}^\infty
{(-1)^{f+1}\over f}
\Tr\left[\left({m\Delta_1+\Delta_2\over \Delta_0}\right)^f\right], 
\label{funcexp} 
\end{eqnarray}
where $\Delta_{k}$ is ${\cal O}((gn_{u,d})^{k})$. In the following analyses, 
we focus on the potential up to ${\cal O}((gn_{u,d})^4)$ and the higher-order 
corrections are neglected. This is valid if the typical energy 
scale of the classical backgrounds are much smaller than the compactification 
scale, namely $2\pi R gn_{u,d}\ll 1$, which is consistent with 
phenomenological constraints as argued. In this case, it is sufficient to 
estimate the first few terms of the Taylor-expansion~\eqref{funcexp} as long 
as there is no IR singularity of the propagators in the internal loops. The IR 
singularities, $\Delta_0=0$, generally exist only in the contributions from 
zero-mode loops, and thus one needs to carry out the summation 
in~\eqref{funcexp}. With this in mind, 
we perform a perturbative expansion of the functional determinant up 
to ${\cal O}((gn_{u,d})^4)$ as follows:
\begin{eqnarray}\notag
&&\hspace{-1cm}  \ln\det[\Delta_0+m\Delta_1+\Delta_2]\\[1mm]\notag
&=&\ln\det[\Delta_0]
+\Tr\bigg[\left(\Delta_2\over \Delta_0\right)-{1\over 2}
\left(m^2\Delta_1^2\over \Delta_0^2\right)
-{1\over 2}
\left(\Delta_2^2\over \Delta_0^2\right)
+\left(m^2\Delta_1^2\Delta_2\over \Delta_0^3\right)
-{1\over 4}\left(m^4\Delta_1^4\over \Delta_0^4\right)
\bigg]\\[2.5mm]\notag
&&\hspace{9cm}+(\textrm{IR div.})+{\cal O}((gn_{u,d})^5)\\[1mm]
\notag
&=&\ln\det[\Delta_0]+i\int{d^4p_E\over (2\pi)^4}
\bigg\{{1\over [p_E^2+m^2]}\Tr[\Delta_2]+{m^2\over [p_E^2+m^2]^2}
\Tr[-{1\over 2}\Delta_1^2]\\[1mm]\notag
&&\hspace{1cm}+{1\over [p_E^2+m^2]^2}
\Tr[-{1\over 2}\Delta_2^2]+{m^2\over [p_E^2+m^2]^3}\Tr[\Delta_1^2\Delta_2]
+{m^4\over [p_E^2+m^2]^4}\Tr[-{1\over 4}\Delta_1^4]\bigg\}\\[2mm]
&&\hspace{9cm}+(\textrm{IR div.})+{\cal O}((gn_{u,d})^5),
\label{pert}
\end{eqnarray}
where possible contributions from the IR divergences are implied. 
In the present case, terms with odd order of $\Delta_1$ have odd orbifold 
parity and vanish. To proceed the perturbative calculation, one should reveal 
explicit forms of the operators in the functional determinant of each K-K 
mode. 

\subsubsection{One-loop correction from the vector multiplet}
We start to evaluate the one-loop correction from the 5D vector multiplet. 
As argued, we derive the effective 4D Lagrangian and then the functional 
integral is performed with the perturbative expansion~\eqref{pert}. 
Let us introduce a gauge fixing function: 
\begin{eqnarray}
G^a&=&{1\over \sqrt{\xi}} \bigg[
\partial^\mu \delta^{ac}A_{\mu}^c-\xi([\partial_y\delta^{ac}
+gf^{abc}a_b]A_y^c+gf^{abc}
\sigma_b\Sigma^c)
\bigg],
\end{eqnarray}
where $\xi$ is a gauge parameter. The gauge fixing is regarded as an extension 
of the well-known 4D $R_\xi$ gauge; gauge fixing terms and ghost Lagrangian 
are thus introduced as 
\begin{eqnarray}\label{laggf}
  {\cal L}_{gf}&=&{-{1\over 2}}G^cG^c,
\qquad 
  {\cal L}_{gh}\;=\;
g\sqrt{\xi}\bar c^a\left[{\delta G^a\over \delta \alpha^c}
\right]c^c.
\end{eqnarray}
Now the quadratic parts of the Lagrangian with respect to the fluctuations are 
written by 
\begin{eqnarray}\notag
{\cal L}_{\rm vec}\big|_{\rm quadratic}&=&  
-{1\over 2}A_\mu^a({\cal D}_{A_\mu}^{\mu\nu})^{ac}A_\nu^c
+\bar c^a({\cal D}_{c})^{ac}c^c
+{1\over 2}A_y^a({\cal D}_{A_y})^{ac}A_y^c
+{1\over 2}\Sigma^a({\cal D}_{\Sigma})^{ac}\Sigma^c
\\
&&
+{i\over 2}\bar \lambda_1^a({\cal D}_\lambda)^{ac}
  \lambda_1^c
+{i\over 2}\bar \lambda_2^a({\cal D}_\lambda)^{ac}
 \lambda_2^c
+{\cal L}_{\rm others}.
\label{flucvec}
\end{eqnarray}
In the above expression, the background dependent operators (${\cal D}$) are 
written by 
\begin{eqnarray}\notag
  ({\cal D}_{A_\mu}^{\mu\nu})^{ac}&=&
\left(-[\square-\partial_y^2]\delta^{ac}+
2gf^{abc}a_b\partial_y+g^2f^{abd}f^{db'c}(a_ba_{b'}+\sigma_b\sigma_{b'})
\right)\eta^{\mu \nu}
+\partial^\mu\partial^\nu(1-\xi^{-1})
\delta^{ac},\\\notag
({\cal D}_{c})^{ac}&=&
-[\square-\xi \partial_y^2]\delta^{ac}+
2g\xi f^{abc}a_b\partial_y+\xi g^2f^{abd}f^{db'c}
( a_ba_{b'}+\sigma_b\sigma_{b'}),
\\
({\cal D}_{A_y})^{ac}&=&
-[\square-\xi \partial_y^2]\delta^{ac}+
2g\xi f^{abc}a_b\partial_y+g^2f^{abd}f^{db'c}
(\xi a_ba_{b'}+\sigma_b\sigma_{b'}),
\\\notag
({\cal D}_{\Sigma})^{ac}&=&
-[\square- \partial_y^2]\delta^{ac}+
2g f^{abc}a_b\partial_y+g^2f^{abd}f^{db'c}( a_ba_{b'}+\xi\sigma_b\sigma_{b'}),
\\\notag
({\cal D}_\lambda)^{ac}&=&\Gamma^M\partial_M\delta^{ac}-gf^{abc}
(i\gamma^5a_b+\sigma_b),
\end{eqnarray}
where $\square$ denotes the four dimensional D'Alambertian operator.  
In~\eqref{flucvec}, mixing between fluctuations $A_y$ and $\Sigma$ 
appears as follows:
\begin{eqnarray}\label{Lothers}
\hspace{-0.8cm}  {\cal L}_{\rm others}&=&
g(\xi-1)A_y^af^{abc}\sigma_b\partial_y\Sigma^c
-g^2A_y^a\left[f^{acd}f^{dbb'}a_b\sigma_{b'}+
f^{abd}f^{db'c}(\sigma_ba_{b'}-\xi a_b\sigma_{b'})\right]\Sigma^c. 
\end{eqnarray}
The above forms are simplified with $\xi=1$ and we adopt this specific choice 
of gauge fixing in the analyses.\footnote{
A factor in~\eqref{Lothers} was misread in ref.~\cite{correcthp}; 
the results of the analysis do not depend on the factor when one focuses  
only on the particular backgrounds of flat directions.}

With the boundary conditions~\eqref{vecbc}, K-K decomposition of the fields 
in the vector multiplets  is written as follows:
{\allowdisplaybreaks
\begin{eqnarray}\notag
  \begin{pmatrix}
  A_\mu^{3}(x,y)\\
  A_y^{1,2}(x,y)\\
  \Sigma^{1,2}(x,y)
  \end{pmatrix}
&=&{1\over \sqrt{2\pi R}}
  \begin{pmatrix}
  (A_\mu^{(0)})^{3}(x)\\
  (A_y^{(0)})^{1,2}(x)\\
  (\Sigma^{(0)})^{1,2}(x)
  \end{pmatrix}
+{1\over \sqrt{\pi R}}\sum_{n=1}^\infty
  \begin{pmatrix}
  (A_\mu^{(n)})^{3}(x)\\
  (A_y^{(n)})^{1,2}(x)\\
  (\Sigma^{(n)})^{1,2}(x)
  \end{pmatrix}
\cos\left({ny\over R}\right),\\\notag
  \begin{pmatrix}
  A_\mu^{1,2}(x,y)\\
  A_y^{3}(x,y)\\
  \Sigma^{3}(x,y)
  \end{pmatrix}
&=&{1\over \sqrt{\pi R}}\sum_{n=1}^\infty  
\begin{pmatrix}
  (A_\mu^{(n)})^{1,2}(x)\\
  (A_y^{(n)})^{3}(x)\\
  (\Sigma^{(n)})^{3}(x)
  \end{pmatrix}
\sin\left({ny\over R}\right),\\\notag
  \begin{pmatrix}
    \eta_1^{1,2}(x,y)\\
    \eta_2^{1,2}(x,y)
  \end{pmatrix}&=&
{1\over \sqrt{\pi R}}
\begin{pmatrix}
  \cos{({\beta y\over R})}&-\sin{({\beta y\over R})}\\
  \sin{({\beta y\over R})}&\cos{({\beta y\over R})}
\end{pmatrix}\\\notag
&&\hspace{2cm}\times 
\left(
    \begin{pmatrix}
0\\
  2^{-1/2}{ (\eta^{(0)}_2)^{1,2}(x)}
    \end{pmatrix}+
\sum_{n=1}^{\infty}
\begin{pmatrix}
(\eta^{(n)}_1)^{1,2}(x)
\sin({ny\over R})\\
(\eta^{(n)}_2)^{1,2}(x)
\cos({ny\over R}) 
\end{pmatrix}    
\right)
,\\\notag
  \begin{pmatrix}
    \eta_1^3(x,y)\\
    \eta_2^3(x,y)
  \end{pmatrix}&=&
{1\over \sqrt{\pi R}}
\begin{pmatrix}
  \cos{({\beta y\over R})}&-\sin{({\beta y\over R})}\\
  \sin{({\beta y\over R})}&\cos{({\beta y\over R})}
\end{pmatrix}\\\notag
&&\hspace{2cm}\times 
\left(
  \begin{pmatrix}
  2^{-1/2} (\eta^{(0)}_1)^3(x)
\\ 0    
  \end{pmatrix}
+\sum_{n=1}^{\infty}
\begin{pmatrix}
  (\eta^{(n)}_1)^3(x)
\cos({ny\over R}) \\
(\eta^{(n)}_2)^3(x)
\sin({ny\over R})
\end{pmatrix}\right). 
\end{eqnarray}}\noindent
Using the K-K decomposition, one can easily carry out the $y$-integral 
in the action from $y=0$ to $y=2\pi R$. Then, the obtained 
4D Lagrangian includes infinite towers of K-K modes in addition to the 
zero modes. 

One can readily derive the contribution to the effective potential from the 
quadratic terms of the 4D effective Lagrangian. The contribution depends 
on the SUSY breaking parameter $\beta$, and is written by  
\begin{eqnarray}\notag
  \delta V_{\rm vec}(\beta)&=&
{-i\over 2\pi R}
{4-2\over 2}
\ln\det[\Delta_0(A_\mu^{(0)})+\Delta_2(A_\mu^{(0)})]\\\notag
&&+{-i\over 2\pi R}
{4-2\over 2}
\sum_{n=1}^\infty
\ln\det[\Delta_0(A_\mu^{(n)})+m_{A_\mu}^{(n)}\Delta_1(A_\mu^{(n)})+
\Delta_2(A_\mu^{(n)})]
\\\notag
&&+{-i\over 2\pi R}
{2\over 2}\ln\det[\Delta_0(Q^{(0)})+\Delta_2(Q^{(0)})]
\\\notag
&&+{-i\over 2\pi R}
{2\over 2}
\sum_{n=1}^\infty
\ln\det[\Delta_0(Q^{(n)})+m_{Q}^{(n)}\Delta_1(Q^{(n)})+\Delta_2(Q^{(n)})]
\\
&&+{i\over 2\pi R}{2\over 2}
\sum_{n=-\infty}^\infty
\ln\det[\Delta_0(\lambda^{(n)})+m_{\lambda}^{(n)}\Delta_1(\lambda^{(n)})+
\Delta_2(\lambda^{(n)})], 
\label{vecdet}
\end{eqnarray}
where $Q^{(n)}\equiv \Sigma^{(n)}+iA_y^{(n)}$, and the fermionic fluctuations
are implied by $\lambda^{(n)}$. The operators $\Delta_i(\varphi)$ denotes the 
contribution from a fluctuation $\varphi$ to $\Delta_i$ and is defined in 
Appendix~\ref{sec:op}. Masses of the bosonic K-K modes, 
$A_\mu^{(n)}$ and $Q^{(n)}$, are the 
same as $m_{A_\mu}^{(n)}=m_Q^{(n)}=n/R$, on 
the other hand, the fermionic fluctuations have mass corrections 
through the Scherk-Schwarz SUSY breaking. Non-zero modes of 
$\eta_1^{(n)}$ and $\eta_2^{(n)}$ mix with each other due to K-K masses 
arising from $y$-derivative in the kinetic terms; they are rearranged into 
mass eigenstates, which are formally written by $n\geq 1$ and $n\leq -1$ 
modes of the masses $m_\lambda^{(n)}=(\beta+n)/R$. Combining the zero-mode 
contributions, all the contributions from fermionic fluctuations are written 
by the summation over $-\infty\leq n\leq \infty$. 

Let us now evaluate the contribution to the effective potential 
up to ${\cal O}((gn_{u,d})^4)$ via the perturbative expansion. 
In~\eqref{vecdet}, we rewrite the functional determinant by the approximate 
form of~\eqref{pert} and obtain the expression as 
{\allowdisplaybreaks
\begin{eqnarray}\notag
\delta V_{\rm vec}(\beta)&\simeq &
{3\over 2\pi R}\sum_{n=-\infty}^\infty \ln\det\left[[\square +(n/R)^2]
-[ \square +{({\beta+n\over R})^2}]\right]\\\notag
&&+{\cal F}_{[1,0,\beta]}(4g^2S)
+{\cal F}_{[{2,1,\beta}]}(-4g^2(S+T))+{\cal F}_{[{2,0,\beta}]}(-4g^4S^2)
\\\notag
&&
+{\cal F}_{[{3,1,\beta}]}(4g^4
(4S^2-D^2+4ST))+{\cal F}_{[{4,2,\beta}]}(-8g^4(S^2+2ST
+T^2))\\\notag
&&+{1\over 2\pi R}
\hat \zeta_{[2,0,0]}
(-g^4D^2)+{1\over 2\pi R}
\int {d^4p_E\over(2\pi)^4 }{1\over [p_E^2]^2}(-2{g^4}D^2)\\\notag
&&+{1\over 2\pi R}
\sum_{f=3}^\infty{(-1)^{f+1}\over f}
\int {d^4p_E\over(2\pi)^4 }{1\over [p_E^2]^f}(2g^2S)^f
\\\notag
&&+{1\over 2\pi R}
\sum_{f=3}^\infty{(-1)^{f+1}\over f}
\int {d^4p_E\over(2\pi)^4 }{1\over [p_E^2]^f}g^{2f}
\left[(S+\sqrt{S^2+3D^2})^f+(S-\sqrt{S^2+3D^2})^f\right]\\
&&-{2\over 2\pi R}
\sum_{f=3}^\infty{(-1)^{f+1}\over f}
\int {d^4p_E\over(2\pi)^4 }{1\over [p_E^2]^f}
\left[(2g^2S)^f\right]\delta_{\beta,0},
\label{cont_vec}
\end{eqnarray}}\noindent
where the Higgs fields $n_{u,d}$ are expressed in terms of the combinations 
$$S\equiv |n_u|^2+|n_d|^2,\qquad T\equiv n_un_d+{\rm h.c.},$$ 
and $D$ in~\eqref{dterm}. These combinations are invariant under the residual 
$U(1)$ gauge transformation. Loop functions including the K-K mode summation 
are defined by 
\begin{eqnarray}
  {\cal F}_{[x,m,\beta]}&=&{1\over 2\pi R}\left[\hat \zeta_{[x,m,0]}-
\hat \zeta_{[x,m,\beta]}\right],\\[2mm]
  \hat \zeta_{[x,m,\beta]}&=&\sum_{n=-\infty}^\infty
\int{d^4p_E\over (2\pi)^4}{({n+\beta\over R})^{2m}\over 
[p_E^2+({n+\beta\over R})^2]^x}, 
\end{eqnarray}
where evaluation of the functions is shown in Appendix~\ref{sec:loop}. The 
last term in~\eqref{cont_vec} is zero except for ${\cal N}=1$ SUSY limit, that 
is $\delta_{\beta\neq 0,0}=0$ and $\delta_{\beta=0,0}=1$, up to 
${\cal O}((gn_{u,d})^4)$. 

Several points should be clarified in~\eqref{cont_vec}. The first line 
contributes to vacuum energy and is independent of the Higgs fields. 
We neglect the irrelevant constants hereafter. The second and third lines 
include the loop functions ${\cal F}_{[x,m,\beta]}$. As shown in 
Appendix~\ref{sec:loop}, the loop integrals with K-K mode summation can be 
divided into two parts: one is an UV divergent integral which respects the 
5D Lorentz invariance and the other is a finite correction which violates the 
invariance. An observation is that the UV divergence does not 
depend on the parameter $\beta$ and thus respects SUSY. On the other hand, 
the finite corrections depend on the SUSY breaking effects. Hence, the 
divergent contributions from bosonic and fermionic fluctuations are canceled 
out in the functions ${\cal F}_{[x,m,\beta]}$, and only UV finite 
contributions remain.\footnote{ 
For $\sigma_{1}=\sigma_{2}=0$, the divergences disappear in each of the 
bosonic and fermionic contributions. The divergences are 
related to $\Sigma^{1,2}$ which are just scalar fields and are not protected 
if SUSY is not there.}

The fourth line also includes loop integrals. The UV divergent contributions 
of the terms proportional to $D^2$ are found. The UV divergence of the first 
term is 5D Lorentz invariant and can be realized as a bulk term. On the other 
hand, the second term explicitly violates the 5D Lorentz symmetry and is 
considered as the divergence localized at the fixed points of the 
orbifold~\cite{GGH}. Again these UV divergences respect SUSY. Such the 
divergent contributions are known to exist in $S^1/Z_2$ exact SUSY theory 
and can be renormalized in a supersymmetric fashion~\cite{Nibb}. We here focus 
on the SUSY breaking contributions; the divergence are simply subtracted and 
a regularized quantity is defined by  
$\Delta V_{\rm vec}\equiv \delta V_{\rm vec}(\beta)-\delta V_{\rm vec}(0)$. 
Thereby, the one-loop corrections are written by finite contributions which 
break both SUSY and 5D Lorentz symmetry.\footnote{
Renormalization procedure may bring higher order contributions~\cite{tlghu} 
and is left for future studies.}

The last three lines of~\eqref{cont_vec} are the contributions from possible 
IR divergent massless propagators of the zero-modes. With a suitable 
regularization of the worse IR behavior, one can extract the IR singularities 
and finite contributions to ${\cal O}((gn_{u,d})^4)$ terms. Notice that the 
IR singularities are also involved in $\hat \zeta_{[2,0,0]}$. Non-vanishing 
VEVs of the Higgs fields provide the physical cutoff of the IR divergences 
and the singularities are canceled out in the final expression as seen below. 

Using the explicit evaluation of the loop functions in the 
Appendix~\ref{sec:loop}, one leads to the one-loop contribution to the 
effective potential. We focus on the case with $\beta\neq 0$, then the 
contribution is written as follows: 
\begin{eqnarray}\notag 
  \Delta V_{\rm vec}
&=&{4g^2
\over 64\pi^5 R^3}
\sum_{w=1}^\infty
{4\sin^2(\pi w\beta)\over w^3}
\left[2S+T\right]\\
&&+{4g^4\over 3\cdot 64\pi^3 R} 
\bigg\{
-19S^2-8ST+2T^2+3D^2
+6S^2\ln\left[{(2\pi R)^22g^2S\over 4\sin^2(\pi\beta)}\right]
\bigg\}, \label{pot_vec}
\end{eqnarray}
where we discard the contribution to the vacuum energy as stated. The last 
logarithmic term includes the Higgs fields, so that a non-zero Higgs VEV 
correctly provides the physical cutoff of the IR divergence of the effective 
potential~\cite{coleman}. 

The result precisely reproduces the evaluation in the flat limit where the 
classical backgrounds are set zero except for a mode corresponding to a flat 
direction of the tree-level potential~\eqref{treesu2susy}, namely the SB 
Higgs. One can choose $\alpha =2\pi gRa_2\neq 0$ and $a_1=\sigma_1=\sigma_2=0$ 
for instance, then the one-loop potential takes 
\begin{eqnarray}
  \Delta V_{\rm vec}\;\to\; {-6\over 64\pi^7R^5}
\bigg\{
-\sum_{w=1}^\infty
{\sin^2(\pi w\beta)\over w^3}\alpha^2+{\alpha^4\over 288}\left[
25-6\ln\left[{\alpha^2\over 4\sin^2(\pi \beta)}
\right]
\right]
\bigg\}. 
\label{flatap_vec}
\end{eqnarray}
This is actually realized as the expansion around $\alpha\ll \beta\lesssim 1$
up to ${\cal O}(\alpha^4)$ of the well-known one-loop 
correction~\cite{potentialSS1,potfor}: 
\begin{eqnarray}
  \Delta V_{\rm vec}^{\rm (flat)}
&=&{-6\over 64\pi^7R^5}\sum_{w=1}^\infty
{\cos(\alpha w)\over w^5}(1-\cos(2\pi w\beta)).
\label{flatpot_vec}
\end{eqnarray}

\subsubsection{One-loop correction from bulk hypermultiplets}
Let us discuss loop corrections from bulk hypermultiplets. 
Since the evaluation is carried out in a similar way as the vector multiplet, 
we briefly summarize the result here. For simplicity, we only introduce the 
fundamental representation in the analyses; generalization to the other 
representation is straightforward. 

Since the hypermultiplets may involve chiral fermions in their zero-modes, 
a gauge anomaly of the residual $U(1)$ gauge symmetry generally 
appears~\cite{5Danom}. The zero-mode contents depend on the parities 
$\eta_{U,P}$ in the boundary conditions~\eqref{hypbc}. Non-zero contribution 
to the anomaly actually emerges from the loop effects of a massless chiral
fermion in the $\eta_U=+1$ hypermultiplet. To evade the awkward case naively, 
we introduce $N_+$ pairs of $(\eta_U,\eta_P)=(+1,+1)$ and 
$(\eta_U,\eta_P)=(+1,-1)$ multiplets in the following analyses. Then, the 
fermion zero-modes always form vector-like pairs and the anomaly is canceled 
out in the contributions from each pair.\footnote{
A divergent tadpole of the residual $U(1)$ D-term, which is known to localize 
at the orbifold fixed point~\cite{5Danom,irges,nilles}, is found in each loop 
contribution from the hypermultiplets. The tadpole contributions from 
$(\eta_U,\eta_P)=(+1,+1)$ and $(\eta_U,\eta_P)=(+1,-1)$ hypermultiplets have 
opposite sign and are canceled out as the anomaly.} 
For $\eta_U=-1$, though there are no massless zero-modes, anomalies 
localized at both the fixed points with opposite sign are induced. 
They are canceled out after the $y$-integration, but cause 
inconsistency in the full 5D theory without the help of some cancellation 
mechanisms. We here simply introduce $N_-$ pairs of $(\eta_U,\eta_P)=(-1,+1)$ 
and $(-1,-1)$ to cancel the anomalies, as the above treatment of $\eta_U=1$. 

With the parity assignments, one can derive K-K expansion 
from~\eqref{hypbc} and 4D effective theory is obtained by carrying out 
the $y$-integration of the 5D Lagrangian~\eqref{lagmat}, where 
the substitution~\eqref{sub} is understood. The quadratic terms of the 
fluctuations in the Lagrangian yields one-loop contributions to the effective 
potential from the bulk hypermultiplets: 
\begin{eqnarray}  \notag
 \delta  V_{\rm hyp}(\beta)&=&
{-i\over 2\pi R}
{2\cdot 2N_+\over 2}
\sum_{n=-\infty}^\infty
\ln\det[\Delta_0(\phi_+^{(n)})+m_{\phi_+}^{(n)}\Delta_1(\phi^{(n)})+
\Delta_2(\phi^{(n)})]
\\\notag
&&+{i\over 2\pi R}
{4N_+\over 2}
\left[\ln\det[\Delta_0(\psi^{(0)})+\Delta_2(\psi_u^{(0)})]
+\ln\det[\Delta_0(\psi^{(0)})+\Delta_2(\psi_d^{(0)})]
\right]
\\\notag
&&+{i\over 2\pi R}
{4\cdot 2N_+\over 2}
\sum_{n=1}^\infty 
 \ln \det[\Delta_0(\psi_+^{(n)})+m_{\psi_+}^{(n)}
\Delta_1(\psi^{(n)})+\Delta_2(\psi^{(n)})]
\\\notag
&&+{-i\over 2\pi R}
{2\cdot 2N_-\over 2}
\sum_{n=-\infty}^\infty
\ln\det[\Delta_0(\phi_-^{(n)})+m_{\phi_-}^{(n)}\Delta_1(\phi^{(n)})+
\Delta_2(\phi^{(n)})]
\\
&&+{i\over 2\pi R}
{4\cdot 2N_-\over 2}
\sum_{n=1}^\infty
\ln\det[\Delta_0(\psi_-^{(n)})+m_{\psi_-}^{(n)}\Delta_1(\psi^{(n)})
+\Delta_2(\psi^{(n)})]
, \label{hypdet}
\end{eqnarray}
where $\phi$ ($\psi$) implies the bosonic (fermionic) fluctuations. 
The operators $\Delta_i(\varphi)$ and the masses $m_\varphi^{(n)}$ in the 
functional determinants are listed in Appendix~\ref{sec:op}.\footnote{
In the present case, eigenvalues of the operators are analytically obtained 
due to the simpleness of the $SU(2)$ fundamental representation. Thus, 
one may try to directly evaluate the functional determinant rather than 
the perturbative calculation~\eqref{pert}. It is, however, difficult 
to carry out the K-K mode resummation.} 
Note that, in contrast to the vector multiplet, bosonic fluctuations have 
the Scherk-Schwarz SUSY breaking masses in the hypermultiplets. Non-zero 
modes of $\phi_1$ and $\phi_2$ mix with each other, and then the 
contributions from the 
bosonic fluctuations are collected into the summation over 
$-\infty\leq n\leq \infty$. 

The functional determinants are evaluated in the same way as the 
case of the vector multiplet. With a non-zero value of $\beta$, the one-loop 
contribution up to ${\cal O}((gn_{u,d})^4)$ is obtained as  
\begin{eqnarray}\notag
  \Delta V_{\rm hyp}&=&
{-2N_+g^2\over 64\pi^5 R^3}\sum_{w=1}^\infty
{4\sin^2(\pi w\beta)\over w^3}
[2S+T]\\[1mm]\notag
&&+{-2N_+g^4\over 12\cdot 64\pi^3 R}\bigg\{
-19S^2-8ST+2T^2+6S^2\ln\left[
{(2\pi R)^2g^2S/2\over 4\sin^2(\pi\beta)}
\right]
\bigg\}\\[1mm]
&&+{-2N_-g^2\over 64\pi^5 R^3}
\sum_{w=1}^\infty{4(-1)^w \sin^2(\pi w \beta)\over w^3}(2S+T)
+{N_-g^4S^2\over 64\pi^3 R}\ln\left[\cos^2(\pi \beta)\right].
\label{pot_hyp}
\end{eqnarray}
As the case with vector multiplet, the flat limit is taken as 
\begin{eqnarray}\notag
  \Delta V_{\rm hyp}&\to&{6\cdot 2N_+\over 64\pi^7 R^5}
\left[-
\sum_{w=1}^\infty
{\sin^2(\pi w\beta)\over w^3}
(\alpha/2)^2
+{(\alpha/2)^4\over 288}
\left\{
25-6\ln \left[{({\alpha/ 2})^2\over 4\sin^2(\pi\beta)}\right]
\right\}
\right]\\[1mm]
&&+{6\cdot 2N_-\over 64\pi^7 R^5}
\left[
-\sum_{w=1}^\infty{(-1)^w\sin^2(\pi w\beta)\over w^3}(\alpha/2)^2+
{(\alpha/2)^4\over 48}\ln\left[\cos^2(\pi \beta)\right]
\right]. 
\label{flatap_hyp}
\end{eqnarray}
This corresponds to the expansion of the known forms of the following 
potential around $\alpha\ll \beta\lesssim 1$ up to 
${\cal O}(\alpha^4)$~\cite{potentialSS1,potfor}:
\begin{eqnarray}\notag
  \Delta V_{\rm hyp}^{\rm (flat)}&=&{6\cdot 2N_+\over 64\pi^7 R^5}
\sum_{w=1}^\infty
{\cos(\alpha w/2)\over w^5}(1-\cos(2\pi w\beta))\\
&&+  {6\cdot 2N_-\over 64\pi^7 R^5}\sum_{w=1}^\infty
{\cos((\alpha/2-\pi)w)\over w^5}(1-\cos(2\pi w\beta)).
\label{flatpot_hyp}
\end{eqnarray}

\subsection{One-loop Higgs potential and mass spectrum}
We examined the bulk loop contributions to the effective potential in the 
previous subsection. Then, the approximate forms of the 
potential~\eqref{pot_vec} and~\eqref{pot_hyp}, which involve all the 
background of the scalar zero-modes, are explicitly obtained up 
to ${\cal O}((gn_{u,d})^4)$. The potential should be regarded as low-energy 
effective Higgs potential. In this subsection, we proceed to study the 
vacuum and mass spectrum using the effective potential. 

\begin{figure}[]
\begin{center}
  \includegraphics[width=6cm,clip]{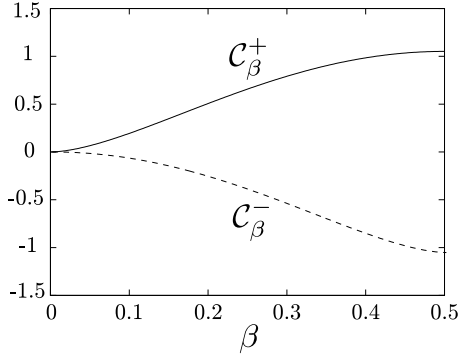}
\end{center}
\caption{
The values of ${\cal C}_\beta^+$ and ${\cal C}_\beta^-$ as the functions 
of $\beta$. 
\bigskip}
\label{fig:sumbeta} 
\end{figure}

In order to make the discussion clear, we introduce the 4D normalization as 
\begin{eqnarray}\notag
 g_c\;=\;gL^{-1/2},\qquad   h_u\;=\;n_uL^{1/2},\qquad  h_d\;=\;n_dL^{1/2},  \\\notag
V_h^0\;=\;LV_{\rm tree}  ,\qquad 
 V_h^{1}\;=\;L(\Delta V_{\rm vec}+\Delta V_{\rm hyp}),
\end{eqnarray}
where $L=2\pi R$, $g_c$ is 4D dimensionless gauge coupling and $h_{u,d}$ 
are up- and down-type Higgs scalars in the canonical normalization. 
From~\eqref{treesu2susy}, \eqref{pot_vec} and~\eqref{pot_hyp}, the Higgs 
potential is derived as follows: 
{\allowdisplaybreaks
\begin{eqnarray}\notag
  V_h^0&=&{g_c^2\over 2}(|h_u|^2-|h_d|^2), \\[2mm]\notag
  V_h^1&=&
{2g_c^2\over \pi^2}{1\over (2\pi R)^2}
\left[\left(1-{N_+\over 2}\right){\cal C}_\beta^+
  -{N_-\over 2}
{\cal C}_\beta^-
\right]
\left[
2(|h_u|^2+|h_d|^2)+(h_u h_d+{\rm h.c.})
\right]\\\notag
&&+
{g_c^4\over 24\pi^2}
\bigg\{
-19(|h_u|^2+|h_d|^2)^2-8(|h_u|^2+|h_d|^2)(h_uh_d+{\rm h.c.})
+2(h_uh_d+{\rm h.c.})^2\\\notag
&&\hspace{2cm}
+3(|h_u|^2-|h_d|^2)^2
+6(|h_u|^2+|h_d|^2)^2
\ln\left[{(|h_u|^2+|h_d|^2)\over {\cal C}_{\beta}^\ell}\right]
\bigg\}\\\notag
&&-{N_+\over 8}{g_c^4\over 24\pi^2}
\bigg\{
-19(|h_u|^2+|h_d|^2)^2-8(|h_u|^2+|h_d|^2)(h_uh_d+{\rm h.c.})
+2(h_uh_d+{\rm h.c.})^2\\\notag
&&\hspace{2cm}+6(|h_u|^2+|h_d|^2)^2
\ln\left[{(|h_u|^2+|h_d|^2)\over 4{\cal C}_{\beta}^\ell}\right]
\bigg\}\\
&&+{6N_-\over 8}{g_c^4\over 24\pi^2} \ln[\cos^2(\pi\beta)] (|h_u|^2+|h_d|^2)^2
\label{vone},
\end{eqnarray}
where 
\begin{eqnarray}\notag
{\cal C}_\beta^+\;=\;
\sum_{w=1}^\infty
{\sin^2(\pi w\beta)\over w^3},\qquad 
{\cal C}_\beta^-\;= \;
\sum_{w=1}^\infty
{(-1)^w\sin^2(\pi w\beta)\over w^3},\qquad 
{\cal C}_\beta^\ell\;=\;{2\sin^2{(\pi\beta)}\over (2\pi R)^2 g_c^2}.
\end{eqnarray}}\noindent
The numerical factors ${\cal C}_\beta^+$ and ${\cal C}_\beta^-$ are displayed 
in fig.~\ref{fig:sumbeta} as the functions of $\beta$. 

The effective potential~\eqref{vone} is invariant under the replacement 
between $h_u$ and $h_d$. Therefore, the field space which satisfies $h_u=h_d$ 
always becomes stationary against the variation corresponding to the mode of 
the non-flat direction, $h_u\neq h_d$. We thus focus on minima along the 
flat direction in the present analyses; if there is no tachyonic mode around 
the minima, the stability of the minima is locally ensured. The Higgs scalars 
are expanded around the minimum as 
\begin{eqnarray}
  h_u&=&{1\over \sqrt{2}}(v_h+r_u+i\pi_u),\qquad 
  h_d\;=\;{1\over \sqrt{2}}(v_h+r_d+i\pi_d),
\label{expand}
\end{eqnarray}
where $v_h$ is the Higgs VEV taken to be real positive, and $r_{u,d}$ 
and $\pi_{u,d}$ are real scalars. The VEV $v_h$ represents the Wilson line 
degree of freedom: $\alpha=\sqrt{2}g_cLv_h$. Thus, the value of $v_h$ is 
dynamically determined through the effective potential~\eqref{vone}, as seen 
below. With a non-trivial value of $v_h$, the residual $U(1)$ gauge symmetry 
is broken and the zero-mode of $A_\mu^3$ acquires the mass of 
$M_W^2=(2g_cv_h)^2$.

We now focus on the physical Higgs mass spectrum. At the tree-level, the four 
real scalars $r_{u,d}$ and $\pi_{u,d}$ are rearranged as the mass eigenstates: 
\begin{eqnarray}\notag
  h&=&{1\over \sqrt{2}}(r_u+r_d),\qquad   H\;=\;{1\over \sqrt{2}}(-r_u+r_d),\\
  A&=&{1\over \sqrt{2}}(\pi_u+\pi_d),\qquad   
  G\;=\;{1\over \sqrt{2}}(-\pi_u+\pi_d),\label{eigen}
\end{eqnarray}
where $h$ and $H$ are the CP-even Higgs scalars, $A$ is the CP-odd Higgs
and $G$ is the mode eaten by the massive gauge boson. In terms of 5D 
language, $h$, $H$, $A$ and $G$ correspond to the zero-modes of $a_2$, 
$\sigma_1$, $\sigma_2$ and $a_1$, respectively. Using~\eqref{expand} in 
$V_h^0$, we observe that only $H$ have the non-zero mass as $m_H^2=M_W^2$ 
and the others are massless at the tree-level. 

Since the one-loop potential lifts up both the flat and non-flat directions,
all the Higgs masses are corrected. As long as the Higgs VEV lies along the 
flat direction, even at the one-loop level, mass eigenstates of the Higgs 
fields are same as~\eqref{eigen}. Putting the expansion~\eqref{expand} 
into $V_h^1$, the one-loop mass corrections are defined by 
\begin{eqnarray}\notag
  \delta m_\phi^2&\equiv&
 {\partial^2 V_h^1\over \partial \phi^2}\bigg|_{r,\pi \to 0},
\end{eqnarray}
where $\phi=\{h,H,A,G\}$. The mass corrections to the physical Higgs modes 
are found to be 
\begin{eqnarray}\notag
  \delta m_h^2&=&{3g_c^2 M_W^2\over 2\pi^2\alpha^2}
\bigg[
(2-N_+){\cal C}_\beta^+
-{N_-}
{\cal C}_\beta^-
+{\alpha^2\over 96}\bigg\{
9N_+-72
+{3N_-}\ln\left[\cos^2(\pi \beta)\right]
\\
&&\hspace{4.2cm}\notag
+24\ln\left[{\alpha^2\over 4\sin^2(\pi\beta)}\right]
-3N_+\ln\left[{\alpha^2\over 16\sin^2(\pi\beta)}\right]
\bigg\}
\bigg],\\[2mm]\notag
  \delta m_H^2&=&{g_c^2 M_W^2\over 2\pi^2\alpha^2}
\bigg[
(2-N_+){\cal C}_\beta^+
-{N_-}
{\cal C}_\beta^-
+{\alpha^2\over 96}\bigg\{
9N_+-48+{3N_-}\ln\left[\cos^2(\pi \beta)\right]
\\
&&\hspace{4.2cm}\label{mco_1}
+24\ln\left[{\alpha^2\over 4\sin^2(\pi\beta)}\right]
-3N_+\ln\left[{\alpha^2\over 16\sin^2(\pi\beta)}\right]
\bigg\}
\bigg],\\[2mm]\notag
  \delta m_A^2&=&{g_c^2 M_W^2\over 2\pi^2\alpha^2}
\bigg[
(2-N_+){\cal C}_\beta^+
-{N_-}
{\cal C}_\beta^-
+{\alpha^2\over 96}\bigg\{
9N_+-72+{3N_-}\ln\left[\cos^2(\pi \beta)\right]
\\
&&\hspace{4.2cm}\notag
+24\ln\left[{\alpha^2\over 4\sin^2(\pi\beta)}\right]
-3N_+\ln\left[{\alpha^2\over 16\sin^2(\pi\beta)}\right]
\bigg\}
\bigg],
\end{eqnarray}
where we use the notation $\alpha=\sqrt{2}g_cLv_h$. A simple relation 
$\delta m_A^2=\delta m_h^2/3$ is observed. In the present case, the 
CP-odd Higgs $A$ always becomes lighter than the SB Higgs $h$; this may be 
affected by the introduction of matter multiplets in larger representations.

It should also be mentioned that the terms proportional to ${\cal C}_\beta^+$ 
and ${\cal C}_\beta^-$ appear as the particular combination, 
$(2-N_+){\cal C}_\beta^+-{N_-}{\cal C}_\beta^-$, in all the mass 
corrections~\eqref{mco_1}, up to the overall factors. These contributions 
come from ${\cal O}((g_ch_{u,d})^2)$ part of the effective 
potential~\eqref{vone}. As argued below, in order to obtain realistic 
vacua ($\alpha\ll 1$), coefficient of the quadratic part of the potential 
should be suppressed. This requires 
$(2-N_+){\cal C}_\beta^+-{N_-}{\cal C}_\beta^-\ll 1$ in the SB 
Higgs mass $\delta m_h^2$. Such the cancellation occurs simultaneously in 
$\delta m_A^2$ and $\delta m_H^2$ in the present analyses. 
The situation however may not be generally realized; if one adds matter 
multiplets in larger representations or considers models with larger gauge 
groups, then the cancellation may no longer occur in mass corrections of 
some of the non-SB Higgs scalars. In such cases, some of the 
non-SB Higgs masses 
could be enhanced compared to the SB Higgs mass through corrections from 
${\cal O}((g_ch_{u,d})^2)$ part of the potential. 

One can read the stationary condition of the potential from $\delta m_G^2=0$, 
which yields  
\begin{eqnarray}\notag
&&3(2-N_+) 
{\cal C}_\beta^+
-{3N_-}
{\cal C}_\beta^-
+{\alpha^2\over 96}\bigg\{
11N_+-88+{3N_-}\ln\left[\cos^2(\pi \beta)\right]\\
&&\hspace{4cm}
+24\ln\left[{\alpha^2\over 4\sin^2(\pi\beta)}\right]
-3N_+\ln\left[{\alpha^2\over 16\sin^2(\pi\beta)}\right]
\bigg\}\;=\;0,  
\label{sta}
\end{eqnarray}
for $\alpha\neq 0$. 
Using~\eqref{sta}, one can eliminate the logarithmic terms in~\eqref{mco_1} 
and simplify the expressions as 
\begin{eqnarray}\notag 
  \delta m_{h0}^2&=& 
{3g_c^2M_W^2\over 2\pi^2\alpha^2}
\left[
2(N_+-2){\cal C}_\beta^++2N_-{\cal C}_\beta^-
+{\alpha^2\over 96}[
16-2N_+]\right],
\\
 \delta m_{H0}^2&=& 
{~g_c^2M_W^2\over 2\pi^2\alpha^2}
\left[
2(N_+-2){\cal C}_\beta^++2N_-{\cal C}_\beta^-
+{\alpha^2\over 96}[
40-2N_+]\right].\label{mco_2}
\end{eqnarray}
These expressions are only valid at the minimum of the one-loop effective 
potential~\eqref{vone}; it is implied by the subscripts $0$ in the left 
hand side in~\eqref{mco_2}. From the expressions of~\eqref{mco_2}, it is 
expected that the ratio between the mass corrections to the SB Higgs $h$ 
and non-SB Higgs $H$ takes order one values. As argued previously, to 
obtain a phenomenologically viable Higgs mass spectrum, the one-loop mass 
correction to the SB Higgs should not be suppressed more than the weak scale. 
Thus, we expect that the correction to the tree-massive Higgs $H$ is not 
negligible compared to the tree-level mass, as discussed in 
Section~\ref{sec:mass}. 

Let us study the mass corrections to Higgs scalars with the dynamically 
determined VEV of the flat direction $\alpha$. The flat limit potential 
is obtained by the substitution $h_u=h_d\to v_h/\sqrt{2}=\alpha/(2g_cL)$ 
in~\eqref{vone}:
\begin{eqnarray*}
\hspace{3cm}  
V_{\rm flat}(\alpha)\;\equiv \;
V_h^{1}(h_u,h_d)\bigg|_{h_u=h_d\to \alpha/(2g_cL)}.
\end{eqnarray*}
If one fix the parameters $\beta$, $N_+$ and $N_-$, then the value of 
$\alpha$ at the minimum of the flat limit potential is dynamically determined; 
we denote the value as $\alpha_0$. A non-trivial value of $\alpha_0$ leads to 
spontaneous breaking of the residual $U(1)$ gauge symmetry. Since the 
approximation used to derive~\eqref{vone} is only valid for 
$\alpha_0\ll \beta\lesssim 1$, we consider such the minima in the analyses. 

It is known that a suppressed value of $\alpha_0$ is obtained when the 
coefficient of the $\alpha^2$ term takes a small negative value and of 
the $\alpha^4$ term is a positive value in the flat limit 
potential~\cite{higgsmass,csaba}. From~\eqref{flatap_vec}, one can see that 
the contributions to the quadratic term from the vector multiplet is 
positive. In addition, from~\eqref{flatap_hyp} and fig.~\ref{fig:sumbeta}, the 
contribution from $\eta_U=+1$ ($\eta_U=-1$) hypermultiplets is realized as 
negative (positive). Thus, $\eta_U=+1$ hypermultiplets play the role to 
decrease the coefficient; for $N_+\geq 3$, the coefficient can be negative. 
Moreover, using the positive contributions from $\eta_U=-1$ hypermultiplets, 
one can obtain a small negative coefficient in the present case. 

In order to increase the one-loop mass correction to the SB Higgs, a large 
positive coefficient of the $\alpha^4$ term in~\eqref{mco_1} is preferred. 
The correction to the coefficient from the vector multiplet is negative and 
from $\eta_U=+1$ ($\eta_U=-1$) hypermultiplets is positive (negative small). 
To obtain a positive value, one must introduce relatively large number of 
$\eta_U=+1$ hypermultiplets since contribution to the $\alpha^4$ term from a 
hypermultiplet in the fundamental representation is suppressed by the factor 
of $1/2^4$ compared to one from the vector multiplet. For the case with  
$N_+\geq 9$, the coefficient takes a positive value and can be enhanced by the 
logarithmic factors in~\eqref{mco_1} for $\alpha_0\ll \beta$~\cite{higgsmass}. 

\begin{figure}[t]
\begin{center}
  \includegraphics[width=7cm,clip]{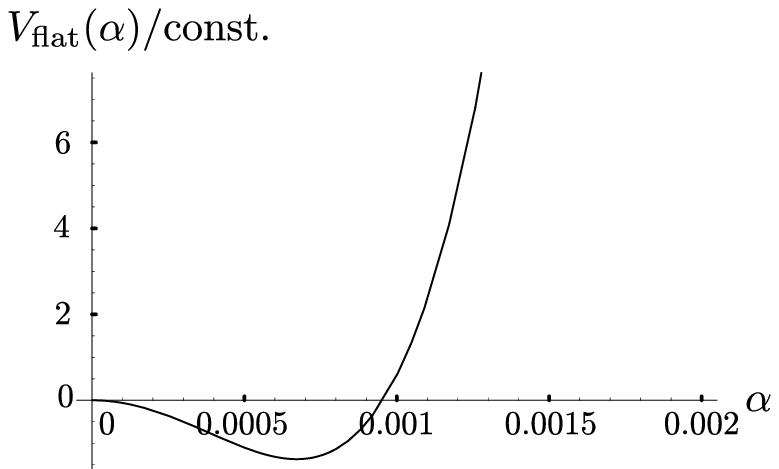}\hspace*{1cm}
  \includegraphics[width=6.8cm,clip]{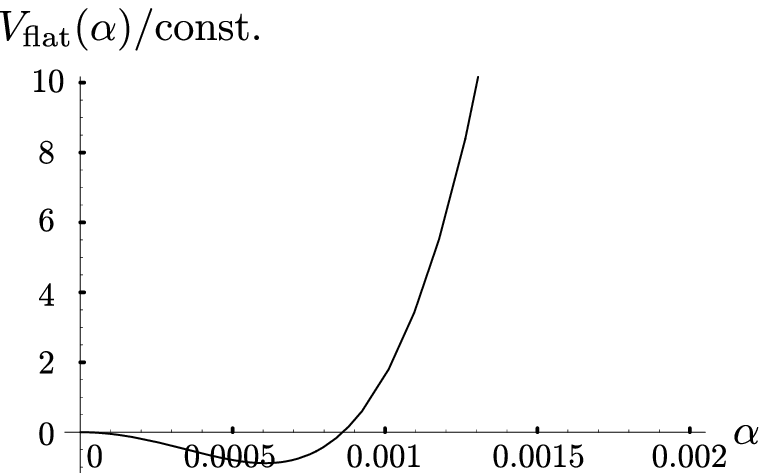}
\end{center}
\caption{
The one-loop Higgs potential along the flat direction $V_{\rm flat}$ 
as the function of $\alpha$, where we take particular normalization of 
the potential. The left figure shows the case with $(N_+,N_-)=(10,15)$ and 
$\beta\simeq 0.211$ and the right one shows the case with $(N_+,N_-)=(30,45)$ 
and $\beta\simeq 0.263$. The potential is minimized with 
$\alpha_0=6.71\times 10^{-4}$ and $\alpha=6.01\times 10^{-4}$ for the 
left and right cases, respectively. 
\bigskip}
\label{fig:pot} 
\end{figure}
As the explicit examples, we show the cases with $(N_+,N_-)=(10,15)$ 
and $(N_+,N_-)=(30,45)$. The number of bulk fields seems to be rather large; 
note that if one introduces hypermultiplets in the adjoint (or larger) 
representation, then a few pieces of bulk fields are sufficient to obtain 
enough heavy Higgs fields~\cite{higgsmass,csaba}. While we don't address the 
issues in the present analyses, introduction of the large representations is 
straightforward. The typical behavior of the potential is shown as the 
function of $\alpha$ in fig.~\ref{fig:pot}, where we take particular 
normalization for the overall scale of the potential. From the figures, 
one can see that $\alpha_0\ll \beta\lesssim 1$ is actually realized; hence, 
the approximation used to derive the effective potential is valid around 
the minima. The numerical evaluation indicates that 
$\alpha_0=6.71\times 10^{-4}$ and $\alpha_0=6.01\times 10^{-4}$ for the 
cases with $(N_+,N_-)=(10,15)$ and $(N_+,N_-)=(30,45)$, respectively. 
The residual $U(1)$ gauge symmetry is broken for both cases. 

\begin{figure}[t]
\begin{center}
  \includegraphics[width=7cm,clip]{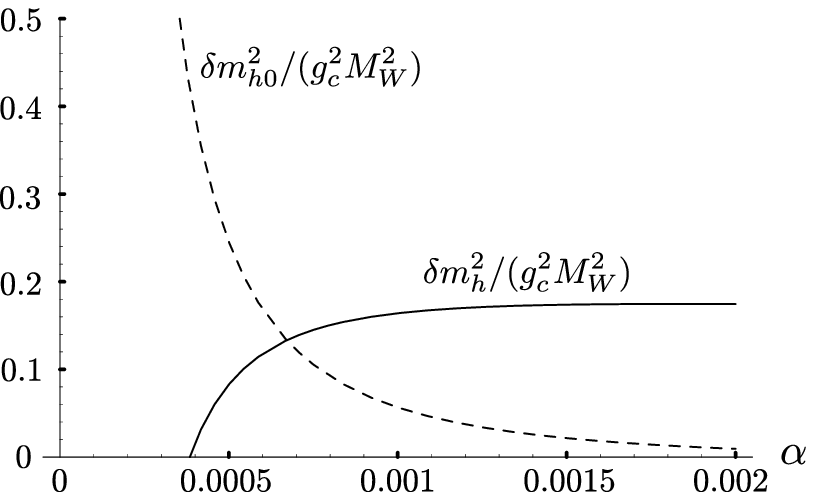}\hspace*{1cm}
  \includegraphics[width=7cm,clip]{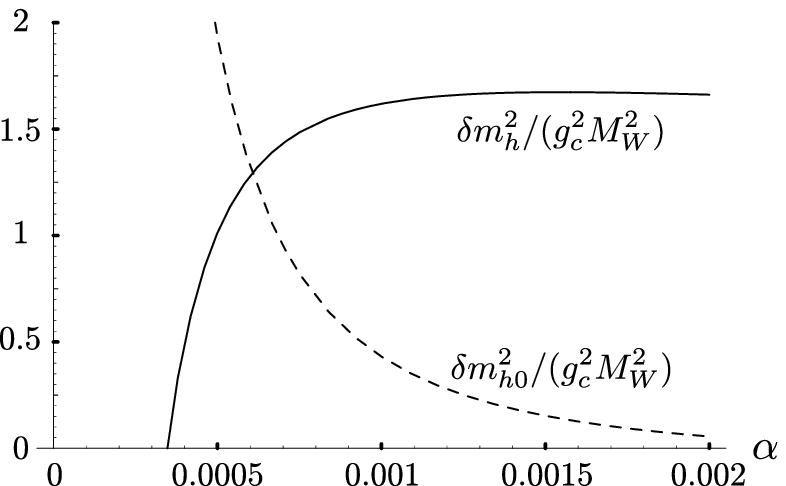}
\end{center}
\caption{
The values of $\delta m_h^2/(g_c^2M_W^2)$ (solid) and 
$\delta m_{h0}^2/(g_c^2M_W^2)$ (dashed) as the functions of $\alpha$. The left 
figure shows the case with $(N_+,N_-)=(10,15)$ and $\beta\simeq 0.211$ and 
the right one shows the case with $(N_+,N_-)=(30,45)$ and $\beta\simeq 0.263$. 
For the illustrative purpose, we draw both the solid and dashed lines in each 
figure: at the minima of the potential, solid and dashed lines indicate 
the same values. We obtain $\delta m_h^2/(g_c^2M_W^2)=0.133$ and 
$\delta m_h^2/(g_c^2M_W^2)=1.32$ for the left and right cases, 
respectively. 
\bigskip}
\label{fig:lighthig} 
\end{figure}
In fig.~\ref{fig:lighthig}, the mass corrections to SB Higgs field $h$ 
are shown as the function of $\alpha$. The solid and dashed lines correspond 
to $\delta m_h^2/(g_c^2M_W^2)$ and $\delta m_{h0}^2/(g_c^2M_W^2)$, 
respectively. For the illustrative purpose, we draw both the lines in each 
figure: at the minima of the potential, both the lines indicate same 
values. We obtain $\delta m_h^2/(g_c^2M_W^2)=0.133$ and 
$\delta m_h^2/(g_c^2M_W^2)=1.32$ 
for the cases with $(N_+,N_-)=(10,15)$ and $(N_+,N_-)=(30,45)$ at the minima, 
respectively. As mentioned before, 
the values of the corrections are straightforwardly increased by incorporating 
more bulk fields and/or large representations~\cite{multi5d,higgsmass,csaba}.

\begin{figure}[t]
\begin{center}
  \includegraphics[width=7cm,clip]{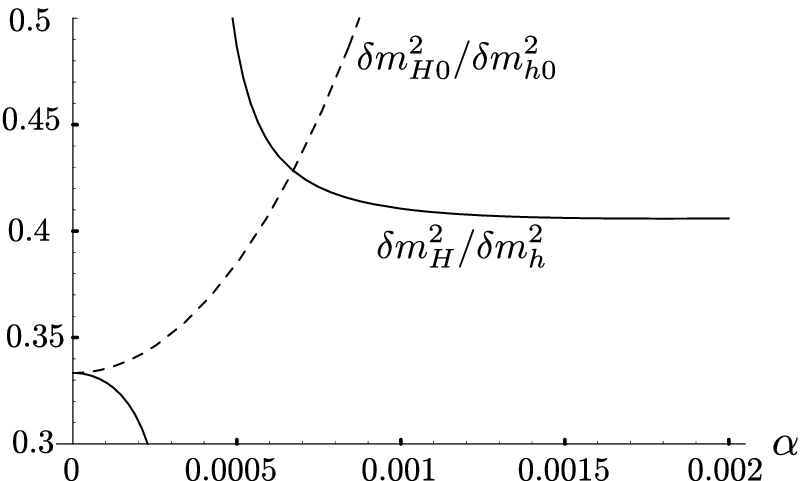}\hspace*{1cm}
  \includegraphics[width=7cm,clip]{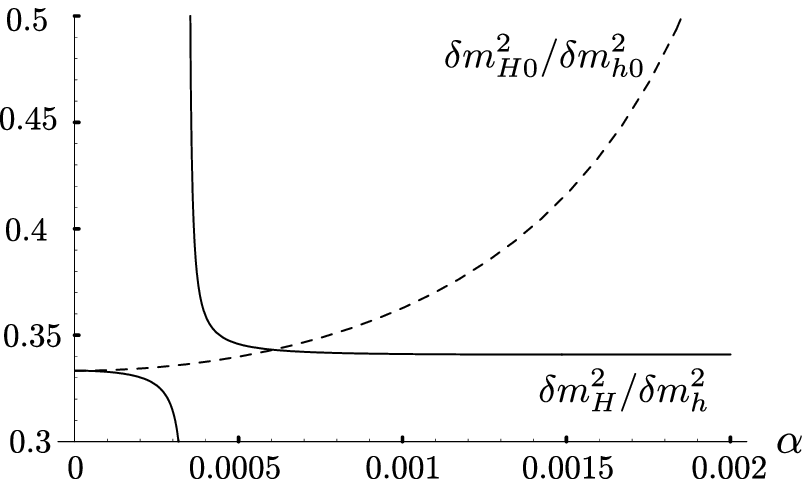}
\end{center}
\caption{
The values of $\delta m_H^2/\delta m_h^2$ (solid) and 
$\delta m_{H0}^2/\delta m_{h0}^2$ (dashed) as the functions of $\alpha$. 
The left figure shows the case with $(N_+,N_-)=(10,15)$ and 
$\beta\simeq 0.211$ and the right one shows the case with $(N_+,N_-)=(30,45)$ 
and $\beta\simeq 0.263$. As in~fig.~\ref{fig:lighthig}, solid and dashed 
lines indicate same values at the minima of the potential;  we obtain 
$\delta m_H^2/\delta m_h^2=0.429$ and for $\delta m_H^2/\delta m_h^2=0.343$ 
the left and right cases, respectively. 
\bigskip}
\label{fig:hevhig} 
\end{figure}
Finally, we show the mass correction to non-SB Higgs $H$. We observe that 
$\alpha$-dependence of the mass correction to $H$ is not so different from 
the case of the SB Higgs $h$. For comparison between the mass corrections to 
$h$ and $H$, we present the ratio between the corrections~in 
fig.~\ref{fig:hevhig}; the solid and dashed lines correspond to 
$\delta m_H^2/\delta m_h^2$ and $\delta m_{H0}^2/\delta m_{h0}^2$, 
respectively. At the point where $\delta m_h^2$ goes to zero, 
$\delta m_H^2/\delta m_h^2$ is strongly enhanced. For a small value of 
$\alpha$, $\delta m_h^2$ and/or $\delta m_H^2$ become negative; thus, the 
solid lines in each figure are disconnected as the functions of $\alpha$. 
The solid and dashed lines are crossed at the minima of the potential; we find 
$\delta m_H^2/\delta m_h^2=0.429$ and $\delta m_H^2/\delta m_h^2=0.343$ for 
the cases with $(N_+,N_-)=(10,15)$ and $(N_+,N_-)=(30,45)$ at the minima, 
respectively. There is no tachyonic mode in the physical Higgs spectrum; 
hence, local stability of the minima is confirmed. The one-loop mass 
correction to the tree-massive mode is not so suppressed rather than the 
tree-level value. The loop correction thus brings important effects not only 
on the tree-massless Higgs scalars, but also on the tree-massive Higgs 
scalar, as expected. 

\section{Summary and perspective}
\label{sec:sum}
In this paper, we considered the 5D $SU(2)$ SUSY model. The model leads 
to the vector-like pair of Higgs scalars at low energy. We explicitly 
analyzed the one-loop corrected multi-Higgs mass spectrum based on the useful 
approximation scheme for the effective potential including all the Higgs 
backgrounds. We focused on the mass correction for the non-SB (tree-massive) 
Higgs and found that both the tree-massless and massive Higgs scalars obtain 
the finite mass collections of similar size from the loop effects.

The results in the analyses implies that the one-loop mass corrections are 
important not only for the tree-massless modes but also for the massive modes 
in several realistic models. For example, in the 5D $SU(3)_c\times SU(3)_W$ 
SUSY model~\cite{multi5d}, two-Higgs doublets appear at a low-energy regime. 
After the electroweak symmetry breaking, physical Higgs spectrum consists of 
two CP-even Higgs scalars $h$ and $H$, a CP-odd Higgs scalar $A$ and a 
charged Higgs scalar $H^{\pm}$. At the tree-level, some of them become 
massive; the one-loop corrections  to their masses are expected to bring 
important effects. It is known that the two Higgs doublets are also predicted 
in the minimal supersymmetric standard model~\cite{inoue}. Except for 
soft SUSY breaking contributions, both models have similar structure of 
tree-level Higgs potential, namely the supersymmetric D-term potential. 
Since the quantum corrections are expected to play the important role to 
raise Higgs masses in both models, comparison between the Higgs spectra is 
interesting. In 6D setup, it is also known to appear two Higgs doublets, and 
the similar Higgs contents to the above models are expected. We expect again 
that the one-loop effects modify the multi-Higgs mass spectrum. 
It is important to estimate the quantum corrections for the Higgs masses in 
view of high-energy experiments. 

Moreover, it is an interesting subject to give detailed study of 
vacuum structure in the above models. In general, minima deviated from the 
flat directions of tree-level potential may be generated through the radiative 
corrections, while such minima are not found in our analyses. 
At the flat direction of the tree-level potential, the ratio 
between VEVs of the two Higgs doublets, so-called $\tan\beta$, is equal to 
one. However, if vacua are shifted to non-flat directions through radiative 
corrections, then $\tan\beta\neq 1$ is realized in the models. The Higgs mass 
spectra may be significantly changed by the shift of the vacua. 

In addition, the perturbative calculation of the effective potential 
developed in this paper is useful for examining the 6D case; approximate 
behavior of the effective potential for a small VEV of the SB Higgs, 
which is not clear even in the flat limit calculation, can be revealed with 
the perturbative expansion of the potential. The examination of the above 
subjects are left to our future studies~\cite{kty2}. 
\bigskip
\subsection*{Acknowledgments}
The authors would like to thank Professor Y.~Hosotani (Osaka Univ.) for 
valuable discussions. K.K. would also thank to K.~Harada (Kyushu Univ.) 
and A.~Watanabe (Kyushu Univ.) for useful discussions. K.T. is partially 
supported by a Grant-in-Aid for Scientific Research (No.~18540275) from 
the Japanese 
Ministry of Education, Science, Sports and Culture and also by the 21st 
Century COE Program at Tohoku University. T.Y. is supported in part by 
The 21st Century COE Program  ``Towards a New Basic Science; Depth and 
Synthesis''. 
\newpage
\appendix

\section{Background dependent operators}
\label{sec:op}
Here we summarize the operators appearing in the functional 
determinants~\eqref{vecdet} and~\eqref{hypdet}. Note that in the determinants, 
using unitary transformations, one can freely change the basis of the 
operators in the representation space. Thus, the following sets of the forms 
are realized in a particular basis. For the contributions from 
vector multiplet in~\eqref{vecdet}, the operators are written as follows:
{\allowdisplaybreaks
\begin{eqnarray*}
\Delta_0(A_\mu^{(0)})&=&\square,\qquad 
\Delta_0(Q^{(0)})\;=\;\square\cdot  {\bf 1}_{2\times 2},\\
\Delta_0(A_\mu^{(n)})&=&\Delta_0(Q^{(n)})\;=\;
[\square+(n/R)^2]\cdot {\bf 1}_{3\times 3},
\qquad m_{A_\mu}^{(n)}\;=\;m_{Q}^{(n)}\;=\;(n/R),
\\
\Delta_0(\lambda^{(n)})&=&[\square+({\beta+n\over R})^2]\cdot
{\bf 1}_{3\times 3},\qquad 
m_\lambda^{(n)}\;=\;(\beta+n)/R,
\\[2mm]
\Delta_1(A_\mu^{(n)})&=&\Delta_1(Q^{(n)})\;=\;\Delta_1(\lambda^{(n)})
\;=\;\sqrt{2}g
\begin{pmatrix}
  0&0&n_u+n_d^*\\
  0&0&n_u^*+n_d\\
  n_u^*+n_d&n_u+n_d^*&0
\end{pmatrix},\\[2mm]
\Delta_2(A_\mu^{(0)})&=&2g^2 (|n_u|^2+|n_d|^2),
\\
\Delta_2(A_\mu^{(n)})&=&
g^2\begin{pmatrix}
  |n_u|^2+|n_d|^2&2n_u n_d^*&0\\
  2n_dn_u^*&|n_u|^2+|n_d|^2&0\\
  0&0&2(|n_u|^2+|n_d|^2)
\end{pmatrix},\\
\Delta_2(Q^{(0)})&=&g^2 
\begin{pmatrix}
  3|n_u|^2-|n_d|^2&2n_un_d^*\\
2n_dn_u^*&3|n_d|^2-|n_u|^2
\end{pmatrix}
,\\
\Delta_2(Q^{(n)})&=&g^2\begin{pmatrix}
  3|n_u|^2-|n_d|^2&2n_u n_d^*&0\\
  2n_dn_u^*&3|n_d|^2-|n_u|^2&0\\
  0&0&2(|n_u|^2+|n_d|^2)
\end{pmatrix},
\\
\Delta_2(\lambda^{(n)})&=&2g^2
\begin{pmatrix}
  |n_d|^2&n_u n_d^*&0\\
  n_u^* n_d&|n_u|^2&0\\
  0&0&|n_u|^2+|n_d|^2
\end{pmatrix}. 
\end{eqnarray*}}\noindent
For the contributions from hypermultiplets 
in~\eqref{hypdet}, the operators are written as follows:
{\allowdisplaybreaks
\begin{eqnarray*}
  \Delta_0(\phi_+^{(n)})&=&[\square+({\beta+n\over R})^2]\cdot
{\bf 1}_{2\times 2},\qquad 
m_{\phi_+}^{(n)}\;=\;(\beta+n)/R,
\\
  \Delta_0(\phi_-^{(n)})&=&[\square+({\beta+n-1/2\over R})^2]\cdot
{\bf 1}_{2\times 2},
\qquad 
m_{\phi_-}^{(n)}\;=\;(\beta+n-1/2)/R,\\
  \Delta_0(\psi^{(0)})&=&\square,\\
  \Delta_0(\psi_+^{(n)})&=&[\square+(n/R)^2]\cdot
{\bf 1}_{2\times 2},
\qquad 
m_{\psi_+}^{(n)}\;=\;n/R,
\\
  \Delta_0(\psi_-^{(n)})&=&[\square+({n-1/2\over R})^2]\cdot
{\bf 1}_{2\times 2},\qquad 
m_{\psi_-}^{(n)}\;=\;(n-1/2)/R,
\\
\Delta_1(\phi^{(n)})&=&\Delta_1(\psi^{(n)})\;=\;g
\begin{pmatrix}
  0&n_u+n_d^*\\
  n_u^*+n_d&0  
\end{pmatrix},\\
\Delta_2(\phi^{(n)})&=&{g^2\over 2}(|n_u|^2+|n_d|^2)\cdot{\bf 1}_{2\times 2},\\
\Delta_2(\psi_{u,d}^{(0)})&=&{g^2}|n_{u,d}|^2,\\
\Delta_2(\psi^{(n)})&=&{g^2}
\begin{pmatrix}
  |n_u|^2&0\\
  0&|n_d|^2
\end{pmatrix}. 
\end{eqnarray*}}\noindent

\section{Loop integrals with K-K mode summation}
\label{sec:loop}
When one evaluates the functional determinant with the perturbative 
expansion~\eqref{pert}, loop integrals with K-K mode summation appear. 
Here the evaluations of the integrals and summations are summarized. 

At first, we introduce the functions, which include 4D loop integral and 
K-K mode summation, as 
\begin{eqnarray}
    \hat \zeta_{[x,m,\gamma]}&=&\lim_{\mu\to 0}
\sum_{n=-\infty}^\infty
\int{d^4p_E\over (2\pi)^4}{({n+\gamma\over R})^{2m}\over 
[p_E^2+({n+\gamma\over R})^2+(\mu/R)^2]^x},
\label{zetahat}
\end{eqnarray}
where $p_E$ is Wick-rotated momentum. A dimensionless parameter $\mu$ is 
introduced in the propagator to regulate IR divergences if any, and we take 
a limit $\mu\to 0$ after the integration. 

For the calculation of the determinant up to ${\cal O}((gn_{u,d})^4)$, 
one needs $\hat \zeta_{[1,0,\gamma]}$, $\hat \zeta_{[2,1,\gamma]}$, 
$\hat \zeta_{[2,0,\gamma]}$, $\hat \zeta_{[3,1,\gamma]}$ 
and $\hat \zeta_{[4,2,\gamma]}$; we consider only them here. Among them, 
$\hat \zeta_{[1,0,\gamma]}$ and $\hat \zeta_{[2,1,\gamma]}$ suffers UV 
divergences from the loop integral, and $\hat \zeta_{[3,1,\gamma]}$ and 
$\hat \zeta_{[4,2,\gamma]}$ suffer IR divergences only. Both the divergences 
are involved in $\hat \zeta_{[2,0,\gamma]}$. In addition, infinite summation 
of the K-K modes also bring the divergence. As shown below, the singularity 
arising from the summation is realized as UV divergence in view of 5D theory. 
Thus, one should evaluate all the functions with a suitable regularization 
both of the UV and IR singularities, even if 4D loop integral does not have 
UV divergence. Using the dimensional regularization, we evaluate UV 
divergences as
\begin{eqnarray}\notag
    \int{d^4p_E\over (2\pi)^4}{1\over 
[p_E^2+\Delta]^x}
&\to&  M_{RG}^{4-d}\int{d^dp_E\over (2\pi)^d}{1\over 
[p_E^2+\Delta]^x}\\
&=&{M_{RG}^{4-d}\over (4\pi)^{d/2}\Gamma(x)}\int_0^\infty
dt t^{(x-d/2)-1}e^{-\Delta t},
\end{eqnarray}
where $d\to 4$ is implied in the expression. We introduce the renormalization 
scale $M_{RG}$ in order to keep the integral having mass dimension $4-2x$. 

To carry out the summation in~\eqref{zetahat}, we use the Poisson resummation 
formulae:
\begin{eqnarray}
    \sum_{n=-\infty}^\infty f(({n+\gamma}))&=&\sum_{w=-\infty}^\infty
e^{2\pi i w\gamma} \int_{-\infty}^\infty dp  f(p)e^{2\pi ip w}.
\end{eqnarray}
Then, in the summation over $w$, only $w=0$ term has local divergence and 
$w\neq 0$ terms are regarded as non-local effects, namely UV finite. Hence, 
we divide $\hat \zeta_{[x,m,\gamma]}$ into two parts as 
\begin{eqnarray}
    \hat \zeta_{[x,m,\gamma]}&=&
  \hat \zeta^{(w=0)}_{[x,m,\gamma]}
+  \hat \zeta^{(w\neq 0)}_{[x,m,\gamma]},
\end{eqnarray}
where $\hat \zeta^{(w=0)}_{[x,m,\gamma]}$ is the $w=0$ term in the 
summation and $\hat \zeta^{(w\neq 0)}_{[x,m,\gamma]}$ is sum of 
the other terms. The locally divergent terms are listed as follows:
\begin{eqnarray}
  \hat \zeta^{(w=0)}_{[1,0,\gamma]}\;=\;{\cal I}_1,\quad
  \hat \zeta^{(w=0)}_{[2,1,\gamma]}\;=\;{{\cal I}_1\over 2},\quad
  \hat \zeta^{(w=0)}_{[2,0,\gamma]}\;=\;{\cal I}_2,\quad
  \hat \zeta^{(w=0)}_{[3,1,\gamma]}\;=\;{{\cal I}_2\over 4},\quad
  \hat \zeta^{(w=0)}_{[4,2,\gamma]}\;=\;{{\cal I}_2\over 8},
\end{eqnarray}
where 
\begin{eqnarray}
  {\cal I}_m&=&2\pi RM_{RG}^{5-(d+1)}\lim_{\mu\to 0}
\int{d^{d+1}p_E\over (2\pi)^{d+1}}{1\over [p_E^2+(\mu/R)^2]^m}.
\end{eqnarray}
One can observe that the divergences arising from 4D loop integral and 
K-K mode summation are written by the 5D Lorentz invariant forms. 
The other non-local UV finite effects are involved in $w\neq 0$ terms, 
and violate 5D Lorentz invariance. 

While $w\neq 0$ terms are UV finite, some of them involve IR divergences, 
as argued. For $\hat \zeta^{(w\neq 0)}_{[1,0,\gamma]}$ and 
$\hat \zeta^{(w\neq 0)}_{[2,1,\gamma]}$, there is no worse IR behavior 
and the integrals can be naively evaluated with $\mu=0$. Then we observe 
\begin{eqnarray}
  \hat \zeta^{(w\neq 0)}_{[1,0,\gamma]}&=&
-\hat \zeta^{(w\neq 0)}_{[2,1,\gamma]}\;=\;
2 {2\pi R\over 64\pi^5 R^3}
\sum_{w=1}^\infty{\cos(2\pi w\gamma)\over w^3}.
\end{eqnarray}
For the others, one should carefully evaluate the following integrals with 
non-zero $\mu$ and $\gamma$:
\begin{eqnarray}
  \hat \zeta_{[2,0,\gamma]}^{(w\neq 0)}&=&\lim_{\mu\to 0}
{2\sqrt{\pi}R\over (4\pi)^{2}}
\sum_{w=1}^\infty \cos{(2\pi w\gamma)}\int_0^\infty dt\ t^{-1/2-1}
e^{-(\mu/R)^2 t}e^{-(\pi w R)^2/t}
,\\
  \hat \zeta_{[3,1,\gamma]}^{(w\neq 0)}&=&\lim_{\mu\to 0}
{-\sqrt{\pi}R\over 2\cdot \Gamma(3)\cdot (4\pi)^{2}}
\sum_{w=1}^\infty \cos{(2\pi w\gamma)}\\\notag
&&\hspace{.3cm}\times \int_0^\infty dt \left[(2\pi w R)^2t^{-3/2-1}
-2t^{-1/2-1}
\right]
e^{-(\mu/R)^2 t}e^{-(\pi w R)^2/t}
,\\
  \hat \zeta_{[4,2,\gamma]}^{(w\neq 0)}&=&\lim_{\mu\to 0}
{\sqrt{\pi}R\over 8\cdot \Gamma(4)\cdot (4\pi)^{2}}
\sum_{w=1}^\infty \cos{(2\pi w\gamma)}\\\notag
&&\hspace{.3cm}\times \int_0^\infty dt \left[
(2\pi w R)^4t^{-5/2-1}
-12(2\pi w R)^2t^{-3/2-1}
+12t^{-1/2-1}
\right]
e^{-(\mu/R)^2 t}e^{-(\pi w R)^2/t}
,
\end{eqnarray}
where $d\to 4$ are taken since they have no UV divergences due to the 
exponential factors including $\omega\neq 0$. Here UV limit corresponds to 
$t\rightarrow 0$ in each $t$-integral. On the other hand, IR limit corresponds 
to $t\to \infty$ and is suitably regularized by non-zero $\mu$. 
After the loop integral, appropriate limit should be taken for $\mu$ and 
$\gamma$. The results are listed for the case with $\gamma=0$ and 
$\gamma\neq 0$ as 
\begin{align}\notag
  \hat \zeta^{(w\neq 0)}_{[2,0,\gamma\neq 0]}&=
  -2  {2\pi R\over 64\pi^3 R}\ln[4\sin^2(\pi \gamma)]+{\cal O}(\mu),&
  \hat \zeta^{(w\neq 0)}_{[2,0,0]}&=
  -2  {2\pi R\over 64\pi^3 R}\ln[(2\pi\mu)^2]+{\cal O}(\mu),\\
  \hat \zeta^{(w\neq 0)}_{[3,1,\gamma\neq  0]}&= 
  {\cal O}(\mu),&
  \hat \zeta^{(w\neq 0)}_{[3,1,  0]}&= 
  -{2\pi R\over 64\pi^3 R}+{\cal O}(\mu),\\\notag
  \hat \zeta^{(w\neq 0)}_{[4,2,\gamma\neq 0]}&=\mathcal{O}(\mu),&
      \hat \zeta^{(w\neq 0)}_{[4,2,0]}&=-{1\over 3}{2\pi R\over 64\pi^3 R}
+\mathcal{O}(\mu), 
\end{align}
where $\mu\to 0$ is understood. Note that $\hat \zeta^{(w\neq 0)}_{[2,0,0]}$ 
includes $\log(\mu)$, which increases with $\mu\to 0$. In the calculation of 
the effective potential, this $\mu$-dependence would be canceled out by the 
other IR divergence. In our example of the calculation in 
Section~\ref{sec:5dsusy}, IR divergences are also involved in the 
resummation of the zero-mode contributions in~\eqref{cont_vec}. Both the 
IR divergences in $\hat \zeta^{(w\neq 0)}_{[2,0,0]}$ and in the resummation 
are canceled out in the one-loop contribution; hence, the effective potential 
does not depend on the artificial parameter $\mu$. 

Finally we define the functions:
\begin{eqnarray}
  {\cal F}_{[x,m,\beta]}&=&{1\over 2\pi R}\left[
\hat \zeta_{[x,m,0]}-\hat \zeta_{[x,m,\beta]}
\right], 
\end{eqnarray}
which appears when contributions from bosonic and fermionic fluctuations are 
summed in SUSY theories. In the functions, all the UV divergences are canceled 
out, and thus there are only the UV finite terms. For a finite value of 
$\beta\neq 0$, they are given as follows: 
\begin{eqnarray}\notag
    {\cal F}_{[1,0,\beta]}&=&
{1\over 64\pi^5 R^3}\sum_{w=1}^\infty
{4\sin^2(\pi w\beta)\over w^3},\\\notag
{\cal F}_{[2,1,\beta]}&=&-{1\over 64\pi^5 R^3}\sum_{w=1}^\infty
{4\sin^2(\pi w\beta)\over w^3},\\
{\cal F}_{[2,0,\beta]}&=&{2\over 64\pi^3 R}
\ln\left[
{4\sin^2(\pi \beta)\over 
({2\pi\mu})^2}
\right]
 ,\\\notag
{\cal F}_{[3,1,\beta]}&=&-{1\over 64\pi^3R},
\\\notag
{\cal F}_{[4,2,\beta]}&=&-{1\over 3\cdot 64\pi^3R},
\end{eqnarray}
where $\mu\to 0$ is understood. For evaluation of the contributions 
from $\eta_U=-1$ fields, it is useful to define the function as
  \begin{eqnarray}
      {\cal F}'_{[x,m,\beta]}&=&{1\over 2\pi R}\left[
\hat \zeta_{[x,m,-1/2]}-\hat \zeta_{[x,m,\beta-1/2]}
\right]. 
\end{eqnarray}
It can be estimated as 
\begin{eqnarray}\notag
    {\cal F}'_{[1,0,\beta]}&=&
{1\over 64\pi^5 R^3}\sum_{w=1}^\infty
{4(-1)^w\sin^2(\pi w\beta)\over w^3},\\
{\cal F}'_{[2,1,\beta]}&=&-{1\over 64\pi^5 R^3}\sum_{w=1}^\infty
{4(-1)^w\sin^2(\pi w\beta)\over w^3},\\\notag
{\cal F}'_{[2,0,\beta]}&=&{2\over 64\pi^3 R}
\ln\left[
\cos^2(\pi \beta)
\right] ,
\end{eqnarray}
and ${\cal F}'_{[3,1,\beta]}={\cal F}'_{[4,2,\beta]}=0$.

\clearpage


\begin{thebibliography}{99}
\bibitem{manton}
  N.~S.~Manton,
  Nucl.\ Phys.\  B {\bf 158} (1979) 141.
\bibitem{fair}
  D.~B.~Fairlie,
  Phys.\ Lett.\  B {\bf 82} (1979) 97.
\bibitem{hosotani}
  Y.~Hosotani,
  Phys.\ Lett.\  B {\bf 126} (1983) 309;
  Annals Phys.\  {\bf 190} (1989) 233.
\bibitem{gaugehiggs1}
  N.~V.~Krasnikov,
  Phys.\ Lett.\  B {\bf 273} (1991) 246;
  H.~Hatanaka, T.~Inami and C.~S.~Lim,
  Mod.\ Phys.\ Lett.\  A {\bf 13} (1998) 2601;
  G.~R.~Dvali, S.~Randjbar-Daemi and R.~Tabbash,
  Phys.\ Rev.\  D {\bf 65} (2002) 064021;
  N.~Arkani-Hamed, A.~G.~Cohen and H.~Georgi,
  Phys.\ Lett.\  B {\bf 513} (2001) 232;
  I.~Antoniadis, K.~Benakli and M.~Quiros,
  New J.\ Phys.\  {\bf 3} (2001) 20.
\bibitem{gaugehiggs2}
  M.~Kubo, C.~S.~Lim and H.~Yamashita,
  Mod.\ Phys.\ Lett.\  A {\bf 17} (2002) 2249;
  L.~J.~Hall, Y.~Nomura and D.~R.~Smith,
  Nucl.\ Phys.\  B {\bf 639} (2002) 307;
  K.~Takenaga,
  Phys.\ Rev.\  D {\bf 64} (2001) 066001;
  Phys.\ Rev.\  D {\bf 66} (2002) 085009;
  C.~Csaki, C.~Grojean and H.~Murayama,
  Phys.\ Rev.\  D {\bf 67}, (2003) 085012;
  G.~Burdman and Y.~Nomura,
  Nucl.\ Phys.\  B {\bf 656} (2003) 3;
  N.~Haba, M.~Harada, Y.~Hosotani and Y.~Kawamura,
  Nucl.\ Phys.\  B {\bf 657}, (2003) 169 
  [Erratum-ibid.\  B {\bf 669}, (2003) 381];
  C.~A.~Scrucca, M.~Serone and L.~Silvestrini,
  Nucl.\ Phys.\  B {\bf 669} (2003) 128;
  I.~Gogoladze, Y.~Mimura, S.~Nandi and K.~Tobe,
  Phys.\ Lett.\  B {\bf 575}, (2003) 66;
  C.~Csaki, C.~Grojean, H.~Murayama, L.~Pilo and J.~Terning,
  Phys.\ Rev.\  D {\bf 69} (2004) 055006.
\bibitem{gaugehiggs3}
  G.~Panico, M.~Serone and A.~Wulzer,
  Nucl.\ Phys.\  B {\bf 739} (2006) 186;
  G.~Panico and M.~Serone,
  JHEP {\bf 0505} (2005) 024;
  N.~Maru and K.~Takenaga,
  Phys.\ Rev.\  D {\bf 72} (2005) 046003;
  M.~Sakamoto and K.~Takenaga,
  Phys.\ Rev.\  D {\bf 75} (2007) 045015;
  Phys.\ Rev.\  D {\bf 76} (2007) 085016;
  N.~Haba, K.~Takenaga and T.~Yamashita,
  Phys.\ Lett.\  B {\bf 605} (2005) 355.
\bibitem{models}
  A.~T.~Davies and A.~McLachlan,
  Phys.\ Lett.\  B {\bf 200} (1988) 305;
  Nucl.\ Phys.\  B {\bf 317} (1989) 237; 
  J.~E.~Hetrick and C.~L.~Ho,
  Phys.\ Rev.\  D {\bf 40} (1989) 4085;
  A.~Higuchi and L.~Parker,
  Phys.\ Rev.\  D {\bf 37} (1988) 2853;
  C.~L.~Ho and Y.~Hosotani,
  Nucl.\ Phys.\  B {\bf 345} (1990) 445;
  A.~McLachlan,
  Nucl.\ Phys.\  B {\bf 338} (1990) 188;
  K.~Takenaga,
  Phys.\ Lett.\  B {\bf 425} (1998) 114. 
\bibitem{warp}
  R.~Contino, Y.~Nomura and A.~Pomarol,
  Nucl.\ Phys.\  B {\bf 671} (2003) 148;
  K.~y.~Oda and A.~Weiler,
  Phys.\ Lett.\  B {\bf 606} (2005) 408;
  K.~Agashe, R.~Contino and A.~Pomarol,
  Nucl.\ Phys.\  B {\bf 719} (2005) 165;
  Y.~Hosotani and M.~Mabe,
  Phys.\ Lett.\  B {\bf 615} (2005) 257;
  Y.~Hosotani, S.~Noda, Y.~Sakamura and S.~Shimasaki,
  Phys.\ Rev.\  D {\bf 73} (2006) 096006;
  Y.~Sakamura and Y.~Hosotani,
  Phys.\ Lett.\  B {\bf 645} (2007) 442;
  Y.~Hosotani and Y.~Sakamura,
  arXiv:hep-ph/0703212; 
  Y.~Sakamura,
  Phys.\ Rev.\  D {\bf 76} (2007) 065002.
\bibitem{multi6d}
  Y.~Hosotani, S.~Noda and K.~Takenaga,
  Phys.\ Rev.\  D {\bf 69} (2004) 125014;
  Phys.\ Lett.\  B {\bf 607} (2005) 276.
\bibitem{multi5d}
  N.~Haba, Y.~Hosotani, Y.~Kawamura and T.~Yamashita,
  Phys.\ Rev.\  D {\bf 70} (2004) 015010;
  N.~Haba and T.~Yamashita,
  JHEP {\bf 0404} (2004) 016.
\bibitem{higgsmass}
  N.~Haba, K.~Takenaga and T.~Yamashita,
  Phys.\ Lett.\  B {\bf 615} (2005) 247.
\bibitem{higgsmass1}
  N.~Maru and K.~Takenaga,
  Phys.\ Lett.\  B {\bf 637} (2006) 287.
\bibitem{tlghu}
  N.~Maru and T.~Yamashita,
  Nucl.\ Phys.\  B {\bf 754}, 127 (2006); 
  Y.~Hosotani, N.~Maru, K.~Takenaga and T.~Yamashita,
  Prog.\ Theor.\ Phys.\ {\bf 118} 1053 (2007). 
\bibitem{correcthp}
  N.~Haba, K.~Takenaga and T.~Yamashita,
  Phys.\ Rev.\  D {\bf 71} (2005) 025006.
\bibitem{nima}
  N.~Arkani-Hamed, T.~Gregoire and J.~G.~Wacker,
  JHEP {\bf 0203} (2002) 055.
 \bibitem{PDG07}
 W.M.~Yao {\it et al.} [Particle Data Group],
 J.~Phys. {\bf G33} (2006) 1.
\bibitem{csaba}
  G.~Cacciapaglia, C.~Csaki and S.~C.~Park,
  JHEP {\bf 0603} (2006) 099.
\bibitem{sohnius}
  M.~F.~Sohnius,
  Phys.\ Rept.\  {\bf 128} (1985) 39; 
  A.~Pomarol and M.~Quiros,
  Phys.\ Lett.\  B {\bf 438} (1998) 255.
\bibitem{bcs}
  M.~Quiros,
  arXiv:hep-ph/0302189.
\bibitem{SS}
  J.~Scherk and J.~H.~Schwarz,
  Phys.\ Lett.\  B {\bf 82} (1979) 60;
  P.~Fayet,
  Phys.\ Lett.\  B {\bf 159} (1985) 121;
  Nucl.\ Phys.\  B {\bf 263} (1986) 649.
\bibitem{SS2}
  R.~Barbieri, L.~J.~Hall and Y.~Nomura,
  Nucl.\ Phys.\  B {\bf 624} (2002) 63.
\bibitem{FIterm}
  P.~Fayet and J.~Iliopoulos,
  Phys.\ Lett.\  B {\bf 51}, (1974) 461.
\bibitem{potentialSS1}
  K.~Takenaga,
  Phys.\ Rev.\  D {\bf 58} (1998) 026004.
\bibitem{potfor}
  N.~Haba and T.~Yamashita,
  JHEP {\bf 0402} (2004) 059.
\bibitem{GGH}
  H.~Georgi, A.~K.~Grant and G.~Hailu,
  Phys.\ Lett.\  B {\bf 506} (2001) 207.
\bibitem{Nibb}
  S.~G.~Nibbelink and M.~Hillenbach,
  Nucl.\ Phys.\  B {\bf 748} (2006) 60.
\bibitem{coleman}
  S.~R.~Coleman and E.~Weinberg,
  Phys.\ Rev.\  D {\bf 7} (1973) 1888.
\bibitem{5Danom}
  N.~Arkani-Hamed, A.~G.~Cohen and H.~Georgi,
  Phys.\ Lett.\  B {\bf 516} (2001) 395;
  C.~A.~Scrucca, M.~Serone, L.~Silvestrini and F.~Zwirner,
  Phys.\ Lett.\  B {\bf 525} (2002) 169.
\bibitem{irges}
  G.~von Gersdorff, N.~Irges and M.~Quiros,
  Phys.\ Lett.\  B {\bf 551} (2003) 351;
  C.~A.~Scrucca, M.~Serone, L.~Silvestrini and A.~Wulzer,
  JHEP {\bf 0402} (2004) 049;
  C.~Biggio and M.~Quiros,
  Nucl.\ Phys.\  B {\bf 703} (2004) 199.
\bibitem{nilles}
  H.~M.~Lee, H.~P.~Nilles and M.~Zucker,
  Nucl.\ Phys.\  B {\bf 680} (2004) 177.
\bibitem{inoue}
  K.~Inoue, A.~Kakuto, H.~Komatsu and S.~Takeshita,
  Prog.~Theor.~Phys. {\bf 67} (1982) 1889.
\bibitem{kty2}
  K.~Kojima, K.~Takenaga and T.~Yamashita, Work in progress.
\end{thebibliography}
\end{document}